\documentclass[10pt,journal,compsoc]{IEEEtran}

\usepackage{multirow}   
\usepackage{makecell}   
\usepackage{enumitem}   
\usepackage{hyperref}   
\usepackage[multiple]{footmisc}     
\usepackage{color, soul}   
\soulregister\cite7
\soulregister\ref7
\soulregister\pageref7

\newcommand{\tabitem}{~~\llap{\textbullet}~~}   

\ifCLASSOPTIONcompsoc
  \usepackage[nocompress]{cite}
\else
  \usepackage{cite}
\fi
\ifCLASSINFOpdf
  \usepackage[pdftex]{graphicx}
  \graphicspath{{../pdf/}{../jpeg/}}
  \DeclareGraphicsExtensions{.pdf,.jpeg,.png}
\else
  \usepackage[dvips]{graphicx}
  \graphicspath{{../eps/}}
  \DeclareGraphicsExtensions{.eps}
\fi
\usepackage{url}

\begin{document}
%

\title{Data Preparation for Software Vulnerability Prediction: A Systematic Literature Review}
%
%
%
%

\author{Roland Croft, Yongzheng Xie, and M. Ali Babar
\IEEEcompsocitemizethanks{\IEEEcompsocthanksitem R. Croft, Y. Xie, and M. A. Babar are affiliated with the Centre for Research on Engineering Software Technologies (CREST), School of Computer Science, University of Adelaide, Adelaide, Australia.\protect\\
E-mail: \{roland.croft, yongzheng.xie, ali.babar\}@adelaide.edu.au
\IEEEcompsocthanksitem R. Croft and M. A. Babar are affiliated with the Cyber Security Cooperative Research Centre, Australia.}
\thanks{Manuscript received...}}

%
%

\markboth{Journal of \LaTeX\ Class Files,~Vol.~14, No.~8, August~2015}%
{Shell \MakeLowercase{\textit{et al.}}: Bare Advanced Demo of IEEEtran.cls for IEEE Computer Society Journals}
%


\IEEEtitleabstractindextext{%
\begin{abstract}
Software Vulnerability Prediction (SVP) is a data-driven technique for software quality assurance that has recently gained considerable attention in the Software Engineering research community. However, the difficulties of preparing Software Vulnerability (SV) related data is considered as the main barrier to industrial adoption of SVP approaches. Given the increasing, but dispersed, literature on this topic, it is needed and timely to systematically select, review, and synthesize the relevant peer-reviewed papers reporting the existing SV data preparation techniques and challenges. We have carried out a Systematic Literature Review (SLR) of SVP research in order to develop a systematized body of knowledge of the data preparation challenges, solutions, and the needed research. Our review of the 61 relevant papers has enabled us to develop a taxonomy of data preparation for SVP related challenges. We have analyzed the identified challenges and available solutions using the proposed taxonomy. Our analysis of the state of the art has enabled us identify the opportunities for future research. This review also provides a set of recommendations for researchers and practitioners of SVP approaches. 
\end{abstract}

}

\maketitle

\IEEEdisplaynontitleabstractindextext

%
\IEEEpeerreviewmaketitle

\ifCLASSOPTIONcompsoc
\IEEEraisesectionheading{\section{Introduction}\label{sec:introduction}}
\else
\section{Introduction}
\label{sec:introduction}
\fi
\IEEEPARstart{S}{oftware} security is a paramount concern due to the continued increase in cybersecurity attacks and exploits that are affecting organizations~\cite{cyberstats2021}. Software security techniques often focus on detecting and preventing Software Vulnerabilities (SVs) that make their way into a software product or deployment pipeline before release~\cite{shahriar2012mitigating}. These SVs are a unique class of software defects that introduce security weaknesses to software and allow for malicious use of products~\cite{shin2013can}. Due to the high importance of removing these security defects, considerable research efforts have been conducted towards their mitigation~\cite{hanif2021rise}. 

Software Vulnerability Prediction (SVP) is a data-driven process for software quality assurance that aims to leverage historical SV knowledge to classify code modules as vulnerable or not. The granularity of the modules can be set as needed, such as file, function, or code snippet. This area of research has recently surged in popularity within the research community~\cite{hanif2021rise}, due to its importance and value. SVP can ensure software security early in development and solve the incapabilities of manually assessing large-scale software systems for potential SVs, which holds inherent value to an organisation.

Like any data-driven process, data preparation serves as one of the most pivotal components for SVP~\cite{pyle1999data}; \textit{garbage in, garbage out.} Consequently, significant efforts need to be expended for data collection and processing~\cite{zheng2018feature}. For SVP, we require examples of both vulnerable and non-vulnerable code for training models. Unfortunately, SV data preparation is not a trivial task~\cite{walden2014predicting}. High-quality SV data is notoriously difficult to obtain due to its natural infrequency~\cite{zimmermann2010searching}, inconsistent reporting~\cite{anwar2020cleaning}, and the unwillingness of organisations to make their sensitive data public~\cite{coulter2020code}. It is widely recognized that data noise can severely impact the quality of an SVP model and eventually negatively impact the validity of the research outcomes~\cite{jimenez2019importance, tantithamthavorn2015impact}. That is why datasets are commonly listed as one of the key challenges for this research area~\cite{lin2020software, hanif2021rise}. These data quality issues, in combination with the extreme data collection effort requirements, have led many to view data as the major barrier to industrial adoption of SVP~\cite{turhan2009relative,morrison2015challenges}. 

However, despite the importance and difficulties of SV data preparation for both industry and academia, there has been relatively little effort allocated to systematically understand the known challenges of data preparation for SVP models and how to address them. Whilst there are several secondary studies that have analyzed SVP research~\cite{lin2020software, zeng2020software, semasaba2020literature, hanif2021rise} and acknowledged the existence of problems with the data preparation, these studies have not focused on thoroughly investigating the data preparation related challenges in SVP research. 

Motivated by a lack of an integrated and comprehensive body of knowledge on this important topic, we aim to highlight the state of the practice of data preparation for SVP and consequently identify the associated SV data preparation challenges and solutions. This knowledge is expected to assist practitioners and researchers in gaining better understanding of the data preparation challenges in SVP and available solutions, in order to support the development and application of more reliable and trustworthy SVP models. In this paper, we empirically examine and synthesize the current practices, challenges and solutions for SVP data preparation through a Systematic Literature Review (SLR). We systematically select 61 peer-reviewed papers on SVP research. We communicate the state of the practice for SVP data preparation techniques, the reported data challenges of these primary studies, and the existing solutions that researchers have used to combat these issues. The main contributions of this research are:
\begin{itemize}
    \item The first, to the best of our knowledge, systematic review aimed at systematically developing an integrated and comprehensive source of information regarding SVP data preparation practices, challenges and solutions, 
    \item A taxonomy of 16 SV data challenges across six themes. This taxonomy can be used to classify data challenges for future SVP research and practice, 
    \item A mapping of the identified solutions onto the data preparation challenges as per the developed taxonomy, 
    \item A set of recommendations on how to overcome the identified data preparation challenges. 
\end{itemize}

The key contributions of this study are expected to help improve the state of the art and the state of the practice of data preparation for SVP models. The findings can raise awareness and understanding about the important challenges of SV data preparation; such understanding will likely assist to avoid the challenges and improve the reliability of SVP models. Furthermore, we provide recommendations to guide the future research on data preparation and data quality assessment for SVP models. Such future research outcomes are expected to ultimately result in more reliable and trustworthy SVP models. The findings from this study can also be leveraged for enhancing the existing or developing new tools for supporting the construction and application of SVP models in general, and the data preparation for SVP models in particular.

The rest of this paper is organized as follows. Section \ref{sec:related} describes the related work and existing SVP reviews. Section \ref{sec:method} presents the methodology we use to conduct our SLR. The findings of our study are presented in Sections \ref{sec:data_prep}, \ref{sec:data_challenges} and \ref{sec:data_solutions}. In Section \ref{sec:recommend}, we provide recommendations for future SV data considerations and research. Finally, in Section \ref{sec:threats}, we state the applicable threats to validity of our findings, and conclude our study in Section \ref{sec:conclusion}. 

\section{Background and Related Work}
\label{sec:related}

SVP is a data-driven process that uses learning-based methods to make predictions, and hence follows the standard Machine Learning (ML) workflow. 

\subsection{Data Preparation}
\begin{figure*}[h]
  \centering
  \includegraphics[width=\linewidth]{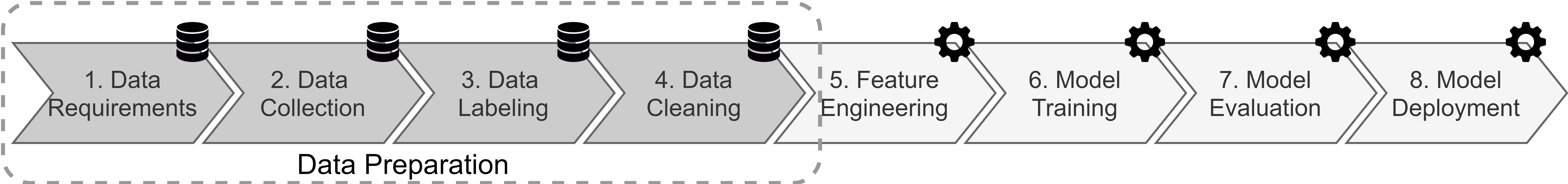}
  \caption{The machine learning workflow. Adapted from Amershi et al.~\cite{amershi2019software}.}
  \label{fig:ml_workflow}
\end{figure*}

Figure \ref{fig:ml_workflow} displays the steps involved in a learning-based workflow~\cite{amershi2019software}. For the purposes of SVP and this study, we consider labeling to occur before cleaning. 

In our analysis of the reviewed papers, we focus on the data-oriented steps (data collection, data labeling and data cleaning) of the ML workflow. The first step of the ML workflow is the model requirements phase, which identifies the necessary requirements and applications of a model. For our study, we consider this step as \textit{data requirements}, as it is necessary to identify the requirements of the data used to build a model, e.g., what kind of data will be used and from where it will be collected. Hence, these data requirements form a necessary preliminary component of data preparation. We collectively define the first four steps of the ML workflow as \textit{data preparation}. 

Practitioners have agreed that the majority of the time taken to construct an ML pipeline is consumed by data preparation~\cite{zheng2018feature}. In the 2019 Appen State of AI survey~\cite{appen2019ai}, it was reported that a majority of practitioners spend upwards of 25\% of their time gathering, cleaning or labeling data. Despite their importance, data preparation processes have rarely been discussed or investigated exclusively~\cite{pyle1999data}.

\subsection{Software Vulnerability Prediction}
\label{sec:rel_svp}
Software Vulnerability Prediction (SVP) models aim to automatically learn SV knowledge and patterns from historical data. This knowledge can be used to make predictions on the presence of SVs. This process was first notably conceptualized in 2007 by Neuhaus et al.~\cite{neuhaus2007predicting}, and has seen continual technical advancement through research efforts~\cite{hanif2021rise}. 

SVP can be considered as an early form of software security quality assurance, as a trained model can make predictions quickly on static code artefacts, without the need for compilation. In this sense, SVP has been compared to static application security testing methods~\cite{croft2021empirical}. Ghaffarian and Shahriari~\cite{ghaffarian2017software} categorized SVP methods into two main approaches: models that do not analyze program syntax and semantics, and models that do. The former utilizes software metrics to describe the code modules of interest, whereas the latter perform directly on source code tokens to perform vulnerable code pattern recognition. Due to the rising popularity of Deep Learning (DL) methods~\cite{hanif2021rise}, researchers have focused more heavily on approaches that analyze program syntax and semantics, through the use of text-based, sequence-based or graph-based source code feature representations~\cite{lin2020software}. 

\begin{figure}[h]
  \centering
  \includegraphics[width=0.9\linewidth]{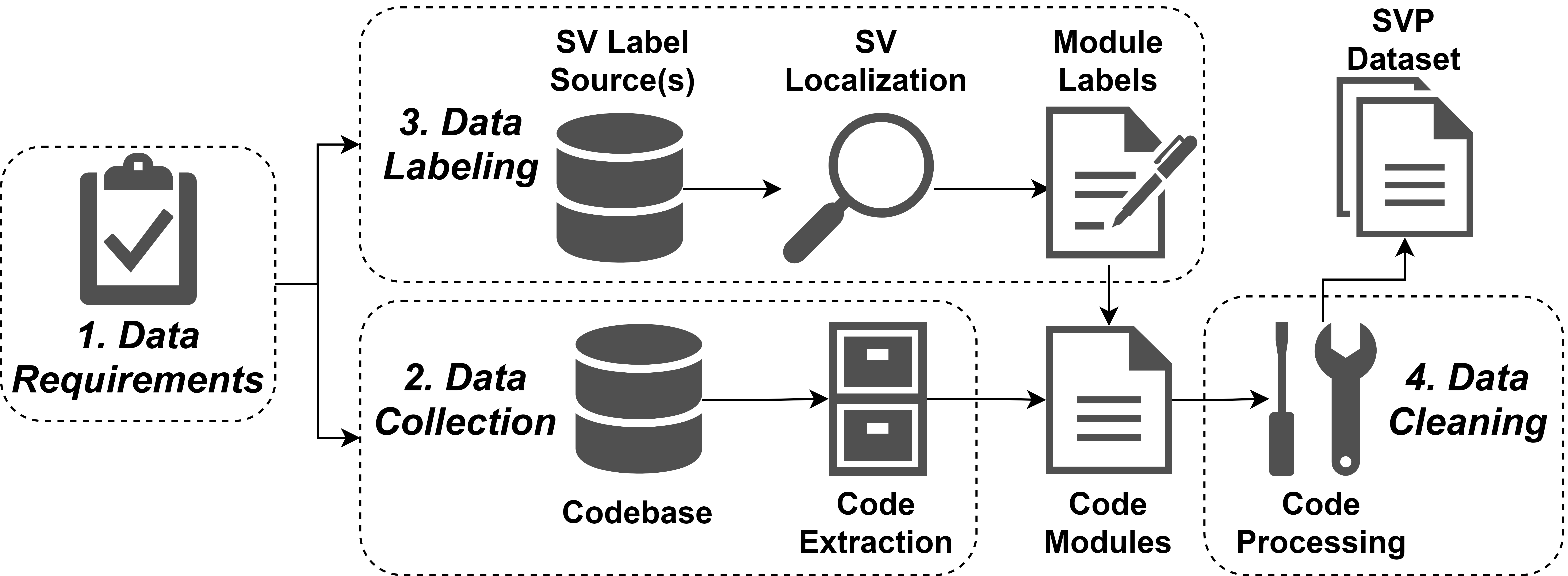}
  \caption{The SVP data preparation pipeline.}
  \label{fig:svp_data_pipeline}
\end{figure}

Data preparation for SVP follows the standard workflow for ML data preparation. SV labels are assigned to the extracted code modules to obtain a labeled SV dataset~\cite{coulter2020code}. The process is heavily dependent on the data sources selected for the codebase and SV labels. Figure \ref{fig:svp_data_pipeline} displays the SVP data preparation pipeline. 

\subsection{Existing SVP Reviews}
\label{sec:rel_studies}
With the increasing popularity of data-driven approaches for software vulnerability analysis and discovery, several researchers have reviewed the published SVP approaches and techniques. We briefly describe the key focus areas of the relevant review studies below. 

Three papers by different groups of researchers, Li and Shao~\cite{li2019survey}, Coulter et al.~\cite{coulter2020code}, and Ghaffarian and Shahriari~\cite{ghaffarian2017software}, reported separate reviews of the literature on the use of machine learning and data mining for software vulnerability discovery and analysis. Coulter et al.~\cite{coulter2020code} provided a more general framework for data-driven cybersecurity tasks, including SVP, whereas Li and Shao~\cite{li2019survey} and Ghaffarian and Shahriari~\cite{ghaffarian2017software} focused on the specific features and approaches. Le et al.~\cite{le2021survey} also conducted a survey of data-driven methods for SV assessment and prioritization, but they did not consider SV discovery. With the success of DL in fields such as image processing, speech recognition, and natural language processing, researchers have been increasingly motivated to apply DL for the SVP domain. Lin et al.~\cite{lin2020software}, Singh and Chatuvedi~\cite{singh2020applying}, and Zeng et al.~\cite{zeng2020software} all conducted an analysis of the deep learning techniques used by researchers for SVP. 

To the best of our knowledge, only two studies have been published that focus on \textit{systematically} reviewing SVP research and knowledge: Semasaba et al.~\cite{semasaba2020literature}, and Hasif et al.~\cite{hanif2021rise}. The former exclusively investigated Deep Learning techniques, whereas the latter provided a wider view of SV detection, including non–learning based techniques. Similar to the previous secondary studies, the analysis of these systematic reviews focused on the models, techniques and features. Furthermore, all existing secondary studies for SVP focused on the model-oriented steps of the ML workflow, particularly features and techniques (steps 5-6 of Figure \ref{fig:ml_workflow}). 

To this extent, there has been little focus on the data-oriented processes. The SV data used to train a model is the most imperative component of this data-driven process. Although most studies have reported data preparation and data quality as significant issues for this research area~\cite{li2019survey, coulter2020code, hanif2021rise, lin2020software, zeng2020software, semasaba2020literature}, they have not performed in-depth analysis of the data quality in SVP research to determine the encountered issues or potential solutions. This knowledge gap fails to provide practitioners and researchers with the specific insights needed to remediate data quality issues. It is vital to gain a better understanding of the quality of data utilized for SVP research; such comprehension is also expected to improve our abilities to better understand how well the SVP approaches work in practice.

Hence, an effort like ours can be of great importance as it not only highlights the critical research gap, but also contributes to the evidence-based body of knowledge of data preparation for SVP models. Whilst existing reviews have yielded important insights into this research domain, our systematic review has been motivated by several unique research questions whose answers have enabled us to provide novel findings and potentially useful insights. The knowledge produced by our SLR can be an important complementary piece to the existing secondary studies for providing a consolidated picture of the published literature on different components of the SVP pipeline.

\section{Research Methodology}
\label{sec:method}

\begin{table*}[t]
  \caption{Research questions addressed in this study.}
  \label{tab:rq}
  \resizebox{\textwidth}{!}{%
  \begin{tabular}{p{5cm}p{12cm}}
    \hline
    \textbf{Research Question (RQ)} & \textbf{Motivation}\\
    \hline
    
    \textbf{RQ1.} What considerations do researchers make for each of the data preparation processes when constructing SVP datasets? 
    & We investigate the SVP data preparation practices and choices reported by researchers to help inform the state of the practice and reasoning. The embodiment of this knowledge helps provide workflow guidance for researchers and practitioners, and assists them to identify the important decisions and reasoning when selecting or building an SV dataset. \\
    \hline
    \textbf{RQ2.} What are the considered challenges and issues for SV data preparation and datasets? 
    & We aim to analyze the reported data challenges to provide an overview of the issues that researchers face when performing SV data preparation or using SV datasets. The results of this RQ will help guide the decisions made by researchers when selecting their data preparation steps, and inform practitioners of the current issues and limitations of the reported empirical SV analysis and prediction. \\
    \hline
    \textbf{RQ3.} How do researchers address dataset issues and preparation challenges? & This RQ builds upon the findings of RQ2 to analyze the remediation techniques that researchers have used to help overcome the aforementioned challenges. Hence we not only provide a categorization of data challenges for SV-related research, but we also map the solutions that researchers have used to address these issues. These findings can help researchers and practitioners overcome data challenges in the future. \\
    
    \hline
\end{tabular}%
}
\end{table*}

\begin{table*}[t]
  \caption{Formulation of the search string.}
  \label{tab:pico}
  \resizebox{\textwidth}{!}{%
  \begin{tabular}{p{1.5cm}p{3cm}p{12.5cm}}
    \hline
    \textbf{Category} & \textbf{Subject} & \textbf{Search Terms}\\
    \hline
    \textbf{\textit{Population}} & Software & \textit{``software'' OR ``code''}\\
    \multirow{2}{*}{\textbf{\textit{Intervention}}} & Machine Learning & \textit{``learn'' OR ``neural network'' OR ``artificial intelligence'' OR ``AI-based'' OR ``predict''}\\
     & Static Application & \textit{NOT (``fuzz'' OR ``test'' OR ``attack'' OR ``adversarial'' OR ``malware'' OR ``description'')}\\
    \textbf{\textit{Comparison}} & - & -\\
    \textbf{\textit{Outcomes}} & Software Vulnerability Prediction & \textit{``vulnerability'' AND (``predict'' OR ``detect'' OR ``classify'' OR ``identify'' OR ``discover'' OR ``uncover'' OR ``locate'')}\\
    \hline
\end{tabular}%
}
\end{table*}

To obtain insights into the SVP data preparation processes, challenges and solutions, we conducted an SLR of SVP literature. Our findings will potentially be useful to both researchers and practitioners for providing guidance for future SV data preparation and in assessing the validity of existing SV datasets. 

To conduct this SLR, we followed the methodological guidance provided by Kitchenham et al.~\cite{kitchenham2004procedures} and Zhang et al.~\cite{zhang2011identifying} to ensure that our assessment of the existing literature was unbiased and repeatable. The research method was conducted in close collaboration by the first two authors, with guidance from the third author. 

\begin{figure}[ht]
  \centering
  \includegraphics[width=0.6\linewidth]{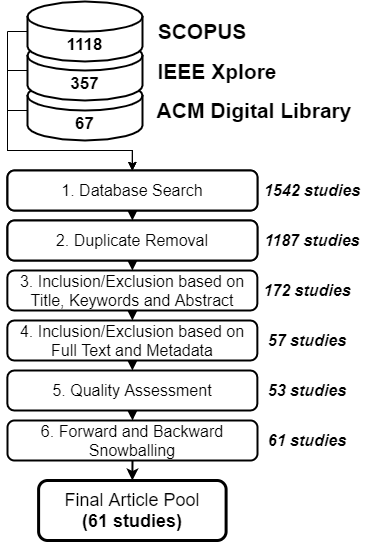}
  \caption{Stages of the study selection process.}
  \label{fig:selection}
\end{figure}

Figure \ref{fig:selection} presents the complete search and study selection workflow, and the number of retrieved papers at each stage. The search and study selection process was conducted in February 2021. We obtained a total of 61 studies from our study selection process, which are presented in the Appendix. 

To guide our analysis, we aimed to address the three Research Questions (RQs) presented in Table \ref{tab:rq}. 

\subsection{Search Strategy}
\label{sec:search}
We began with a search strategy to extract all potentially relevant research papers from academic digital libraries.

To design the search string, we utilized the PICO (Population, Intervention, Comparison, Outcomes) framework~\cite{schardt2007utilization}. The \textit{Comparison} component was not applicable to our review because our goal was not to conduct comparison of software with different interventions. Table \ref{tab:pico} presents the key terms for each PICO component; we formed our search strings through the union (AND) of the PICO components. We altered the search string suitably to match the differences in the search capabilities of each database. When applicable, we matched the relevant keywords in the title, abstract and keywords of the papers, except for exclusion keywords (prefaced with NOT) which were only matched in the title. Wildcard matching was performed to capture different word variants when available; otherwise we defined the term variants manually, e.g., \textit{predict} and \textit{prediction}. When available, we applied additional search filters to match the exclusion criteria defined in Section \ref{sec:selection}.1 (i.e., limiting to English articles or research papers.) Table \ref{tab:pico} only defines the base strings. To find these strings, we consulted papers included in the previous reviews. The full search strings are available in our online appendix\footnote{\url{https://github.com/RolandCroft/SVP_Data_SLR_Appendix/blob/main/Search_Strings.md}}. 

We applied this search string to the two most frequently used academic digital libraries for software engineering, as identified by Zhang et al. [13]: IEEE Xplore, and ACM digital library. We then additionally included SCOPUS as it is the largest academic literature database available [25], which indexes several other smaller academic databases. We did not use other search engines such as Google Scholar due to the amount of noise in the search results and need for subjective stopping conditions. We initially retrieved 1542 studies: 1118 studies from SCOPUS, 357 studies from IEEE Xplore, and 67 studies from ACM Digital Library. We downloaded all retrieved studies and then manually removed duplicates, which reduced the total number of studies to 1187. 

\subsection{Study Selection}
\label{sec:selection}
We sought to select any paper on the topic of Software Vulnerability Prediction (SVP), which we have defined as any model utilising supervised learning-based techniques (ML or DL) for prediction, detection or discovery of an SV in a static code module. We included any study that contributes an SVP model, process or evaluation based on our definition of SVP. 

Our definition of SVP hinges on three major principles: \textit{learning-based}, \textit{vulnerability discovery}, and \textit{static code artefacts}. Firstly, we have defined learning-based as the use of a supervised ML or DL algorithm that can learn from training data to make predictions on a dataset [31]. To this extent, we did not include anomaly detection, unsupervised methods, or studies that focus on pure statistical or correlation analysis. Secondly, the study must have utilized a model that aims to \textit{discover} unknown SVs within a code artefact. This excluded methods that used code clones or similarity detection, as these methods are only able to detect a pre-defined set of SVs and are unable to discover new types. We also did not consider malicious code as SVs. Thirdly, we only included studies that used static code artefacts to make predictions; either source code, code binaries or an intermediary representation. Hence, we excluded any study that requires runtime analysis of the code (e.g., dynamic testing or attack detection). 

\subsubsection{Inclusion/Exclusion Criteria}
The inclusion/exclusion criteria we adopted, displayed in Table \ref{tab:inc_criteria}, were inspired by similar studies~\cite{hosseini2017systematic, li2020systematic}. 

\begin{table}[ht]
  \caption{The inclusion/exclusion criteria.}
  \label{tab:inc_criteria}
  \resizebox{\columnwidth}{!}{%
  \begin{tabular}{p{8cm}}
    \hline
    \textbf{Inclusion Criteria}\\
    \hline
    I1. The study relates to the field of SVP, and informs the practice of Software Engineering. \\
    I2. The study presents a unique SVP process or evaluation. \\
    I3. The study is a full paper longer than six pages. \\
    \hline
    \textbf{Exclusion Criteria}\\
    \hline
    E1. Solely a literature review or survey article. \\
    E2. Non peer-reviewed academic literature. \\
    E3. Academic articles other than conference or journal papers, such as book chapters or dissertations. \\
    E4. Studies not written in English. \\
    E5. Studies whose full-text is unavailable. \\
    E6. Studies published to a venue unrelated to the discipline of Computer Science. \\
    E7. Studies that are published to a journal or conference with a CORE ranking of less than A and H-index less than 40, \textit{and} that have a citation count of less than 20. \\
    \hline
\end{tabular}%
}
\end{table}

To ensure that we obtained a set of high-quality papers, we adopted a venue assessment approach (E7) used by Sabir et al.~\cite{sabir2021machine}. We removed the studies published in low quality venues: venue ranking below A using the CORE ranking system\footnote{\url{http://portal.core.edu.au/conf-ranks/}}\footnote{\url{http://portal.core.edu.au/jnl-ranks/}}, and h-index below 40 as recorded in the Scimago database\footnote{\url{https://www.scimagojr.com/journalrank.php}}. However, the original influential papers of this domain may have been published in low quality venues. Hence, we only excluded a paper based on venue if it had also not been cited frequently (\textless  20 citations). The citation count is obtained through Google Scholar\footnote{\url{https://scholar.google.com.au/}}. The first two authors collaboratively determined suitable thresholds for this criterion through an initial pilot study of 100 papers, to confirm that any papers excluded through this criterion were indeed of lower quality. 

We first excluded 1015 studies using information extracted from the title, abstract and keywords. We then excluded an additional 115 studies after processing the full text and metadata (i.e., venue, article type, citations) to obtain a set of 57 studies. 

\subsubsection{Quality Assessment}
For SLRs, it is vital to assess the quality of primary studies to ensure that we form a proper and fair representation of the research works~\cite{kitchenham2004procedures}. We conducted the assessment process using a quality checklist, and excluded any study that did not pass the checklist. We adopted the quality checklist defined by Hall et al.~\cite{hall2011systematic}, and refined by Hosseini et al.~\cite{hosseini2017systematic} in their SLR of defect prediction models, as the defect prediction process shares similarities with SVP. This resulted in three stages of assessment: the data, the prediction model details, and the evaluation criteria, displayed in Table \ref{tab:quality_assessment}. Although our study only considers data preparation for analysis, we assessed all three criteria to determine the overall quality of the paper. We removed a total of four studies that did not pass the quality assessment criteria.

\begin{table}[ht]
  \caption{The quality checklist.}
  \label{tab:quality_assessment}
  \resizebox{\columnwidth}{!}{%
  \begin{tabular}{p{8cm}}
    \hline
    \textbf{Data Criteria}\\
    \hline
    DC1. The data source must be reported. If a publicly available dataset is used, the name must be reported. \\
    DC2. A description of the data, such as its size, programming language and class distribution, must be provided. \\
    DC3. The process in which the independent variables are extracted from the data as input to the model must be clearly stated. \\
    DC4. The method in which the data is labeled as vulnerable and non-vulnerable must be clearly stated. \\
    \hline
    
    \textbf{Prediction Model Details Criteria}\\
    \hline
    MC1. The output of the model must be clearly defined. \\
    MC2. The granularity of the dependent variable(s) must be reported. \\
    MC3. The machine learning method and approach must be clearly reported. \\
    \hline
    
    \textbf{Evaluation Criteria}\\
    \hline
    EC1. The performance measure of the model must be reported. \\
    EC2. The predictive performance values must be clearly presented in terms of raw performance numbers, means or medians. \\
    \hline
\end{tabular}%
}
\end{table}

\subsubsection{Snowballing}
It is expected that an initial automated search strategy will be unable to identify all relevant studies, as the search string cannot identify obscurely phrased studies, and the digital libraries selected do not exhaustively include all peer-reviewed literature~\cite{wohlin2014guidelines}. Hence, after we conducted initial study selection, we utilized manual search processes, both forward and backward snowballing, to obtain additional relevant studies that were not contained in our selected digital libraries or identified by our automatic search. Forward and backward snowballing identify additional relevant studies from papers that cite or are included in the reference lists of the set of included studies, respectively~\cite{wohlin2014guidelines}. These identified papers were similarly assessed using the inclusion/exclusion criteria and the quality assessment criteria. We included an additional eight papers in the final set through the snowballing process. 

Our final article pool contained \textit{61 studies}; 53 studies which passed the initial selection process and eight additional snowballed papers. The studies are listed in the Appendix. 

\subsection{Overview of the Primary Studies}
\label{sec:bibliometric}

\begin{figure}[ht]
  \centering
  \includegraphics[width=0.9\linewidth]{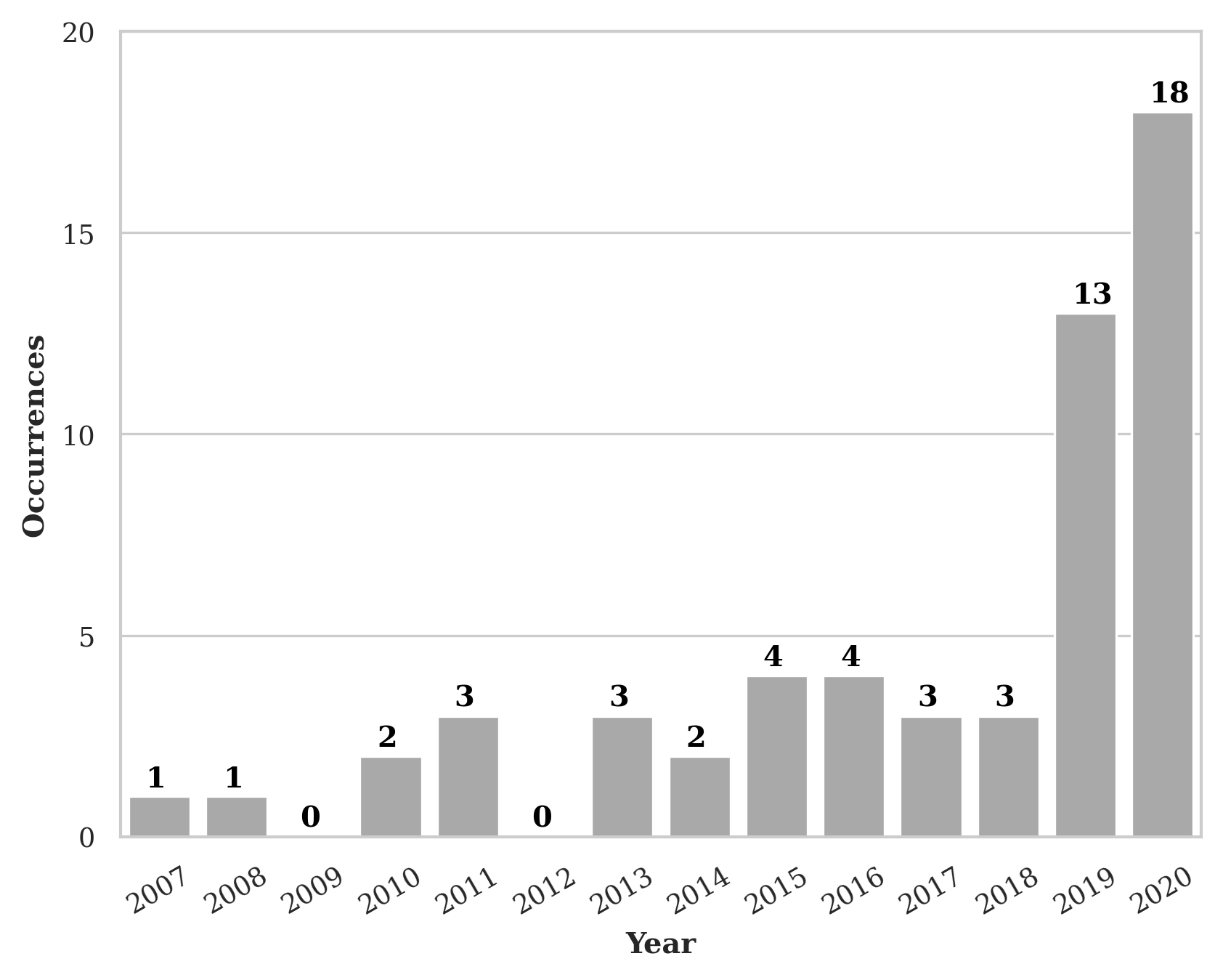}
  \caption{The number of selected primary studies by year.}
  \label{fig:year_trends}
\end{figure}

Figure \ref{fig:year_trends} displays the number of selected SVP papers over the years. We have not reported values for 2021 as this data is incomplete. We observed that this area of research has received exponential popularity within the last two years. This indicates that this is an area undergoing huge growth at the time of this study. Hence, our review contributes important and timely value to this emerging area by synthesizing the current knowledge of the underlying data preparation processes for these empirical studies. 

\subsection{Data Analysis}
\label{sec:analysis}

\subsubsection{Data Extraction}
\label{sec:data_analysis}
We used a data extraction form and the data extraction processes outlined by Garousi and Felderer~\cite{garousi2017experience} and Kitchenham et al.~\cite{kitchenham2004procedures}. Our data extraction form, provided in our online appendix\footnote{\url{https://github.com/RolandCroft/SVP_Data_SLR_Appendix/blob/main/Data_Extraction_Form.pdf}}, consisted of 50 fields describing all data related steps reported by the authors and the details of their dataset(s). These fields consisted of five checkbox questions, 29 multiple choice questions, nine short answer questions, and seven long answer questions. Thirty seven of the 50 fields pertained to the first RQ, and the other 13 collectively related to the latter two RQs. The data extraction form was completed collaboratively using Google Sheets. 

An initial pilot study of 10 papers was conducted collaboratively by the first two authors to help design the form and ensure author agreement~\cite{garousi2017experience}. The first two authors then performed data extraction individually; the paper set was divided in half randomly for each author to complete. After this process was completed, each author reviewed the data extraction outputs of the other author to ensure consistency. Disagreements were resolved through discussion. 

\subsubsection{Data Synthesis}
\label{sec:data_synthesis}
The aim of an SLR is to aggregate information from primary studies~\cite{kitchenham2004procedures}. For RQ1, we qualitatively examined the outputs of our data extraction form to identify and report the major factors relating to each of the four data preparation steps. 

For RQ2 and RQ3, we used thematic analysis to synthesize the data~\cite{braun2006using}. Specifically, this process was used to identify the reported data challenges and solutions. Any discussion in a paper that explicitly had mentioned a challenge pertaining to the data, resolved or unresolved, was coded. To ensure that this qualitative coding was grounded by the data, and not affected by any biases of the data extraction form, we imported the full papers into Nvivo~\cite{Nvivo}, a qualitative data analysis tool, and performed coding on the papers directly. We followed the steps for thematic analysis developed by Braun and Clarke~\cite{braun2006using}: 
\begin{enumerate}
    \item \textit{Familiarizing with data}: The initial familiarization was done through the data extraction phase (Section \ref{sec:data_analysis}) in which the first two authors read each full paper and filled the data extraction form. This familiarized the first two authors with the relevant factors relating to SV data that were discussed in the papers. 
    \item \textit{Generating initial codes}: To generate initial codes, we used \textit{open coding} of the relevant text in the primary studies using Nvivo. The data was broken down into smaller components and labeled using a code~\cite{braun2006using}, where a code is a word or phrase that acts as a label for a selection of meaningful text in the paper. This process was completed iteratively, with the initial codes being revised and merged in later rounds. Each primary study was usually allocated to more than one code or theme, as each paper can discuss multiple SV data challenges and coding was done on small individual components of the papers. 
    \item \textit{Searching for themes}: We reviewed all the codes and sorted them into themes. As data challenges revolve around data quality, we used existing data quality dimensions~\cite{pipino2002data,sidi2012data} to identify potential groupings that the codes might fall under. 
    \item \textit{Reviewing themes, defining and naming themes}: This process involved reviewing all the codes and themes, and revising their allocations. 
    \item \textit{Producing the report}: We present the findings of our thematic analysis in Sections \ref{sec:data_challenges} and \ref{sec:data_solutions}. 
\end{enumerate}

\section{SV Data Preparation Considerations (RQ1)}
\label{sec:data_prep}
\begin{figure}[ht]
  \centering
  \includegraphics[width=0.75\linewidth]{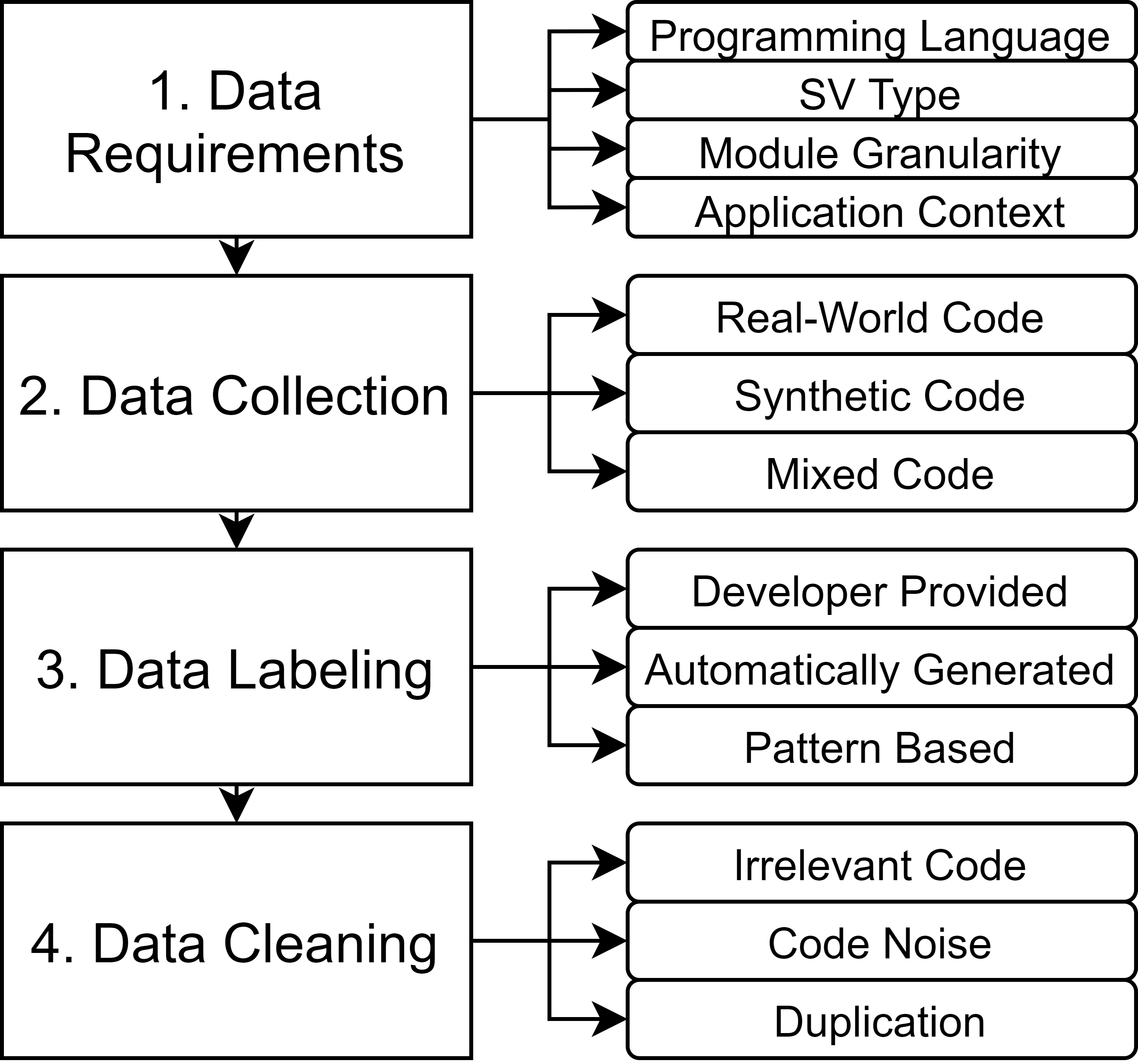}
  \caption{SVP data preparation step considerations.}
  \label{fig:data_considerations}
\end{figure}

We first provide an overview of the considerations that researchers have made when performing SV data preparation processes, which we have identified qualitatively through our data extraction process. This documentation of the considerations helps to inform practitioners and researchers of the state of the practice. Furthermore, it can assist these users to better understand how to construct an SVP dataset, and the important aspects to scrutinize. Figure \ref{fig:data_considerations} displays the main decisions that need to be made for each data preparation step. 

\subsection{Data Requirements}
\begin{figure}[ht]
  \centering
  \includegraphics[width=0.9\linewidth]{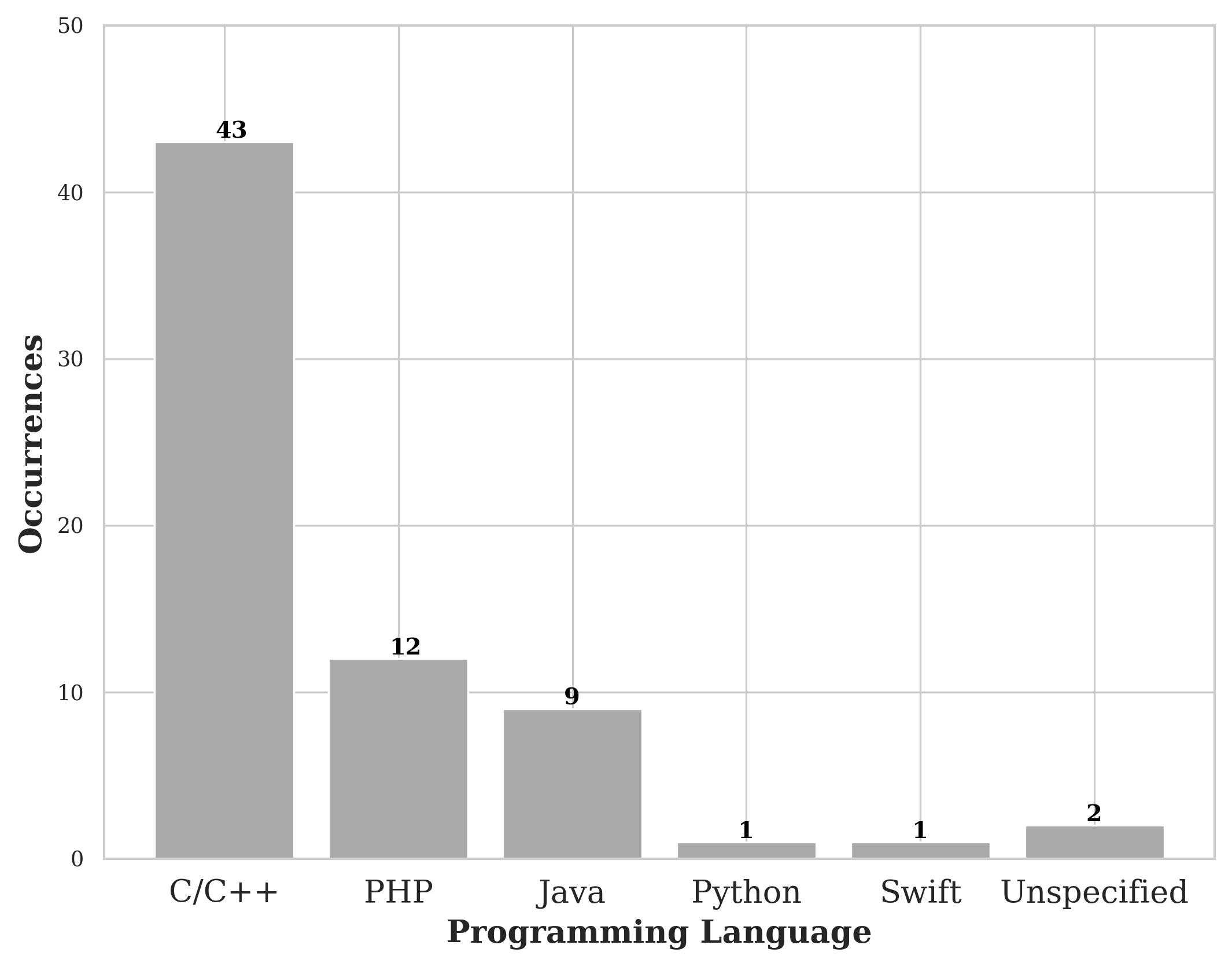}
  \caption{The number of primary studies for each programming language.}
  \label{fig:datasets_in_languages}
\end{figure}

In the data requirements phase, the requirements for the data to achieve the desired model context and capabilities are specified. There are four main components of the data requirements that need to be specified for SVP: 

\textbf{Programming Language(s).}
Researchers are motivated to explore different approaches to mitigate security risks for different languages. As seen in Figure \ref{fig:datasets_in_languages}, C/C++, PHP and Java have been the most commonly investigated languages among the primary studies.

C/C++ has been most frequently chosen for analysis as it has a lower level of abstraction and is commonly used to build security critical applications [P4, P16, P54, P58]. PHP has been commonly used for programming web applications, which are highly susceptible to vulnerabilities and exploits [P1, P6, P25, P41, P50, P55], and hence researchers have aimed to ensure its security. Java has also been commonly chosen as it is overall one of the most popular programming languages [P39, P40, P46]. 

\textbf{Vulnerability type(s).}
Similar to the previous consideration, researchers may target detection capabilities towards certain SV types. However, we observed that the majority of methods are capable of detecting a variety of SVs. Over 45\% of studies (28 out of 61) did not even report the types of SVs present in the data, and researchers were often limited to the SV types present in their SV label source. However, some studies chose to restrict their analysis to more critical SVs of interest. For example, Fidalgo et al. [P6] and Shar and Tan [P25] focused their analysis on SQL injection and cross-site scripting (XSS) as these are common critical web application vulnerability types. Wang et al. [P29] and Ghaffarian and Shahriari [P39] limited their studies to just the CWE Top-25 vulnerabilities\footnote{\url{https://cwe.mitre.org/top25/archive/2021/2021_cwe_top25.html}}. Saccente et al. [P34] identified that a model trained to predict any SV type produces unreliable predictions in comparison to a model trained to predict just one type. However, the impacts of this data requirement decision on model efficacy are otherwise largely under-explored. 

\begin{figure}[ht]
  \centering
  \includegraphics[width=0.9\linewidth]{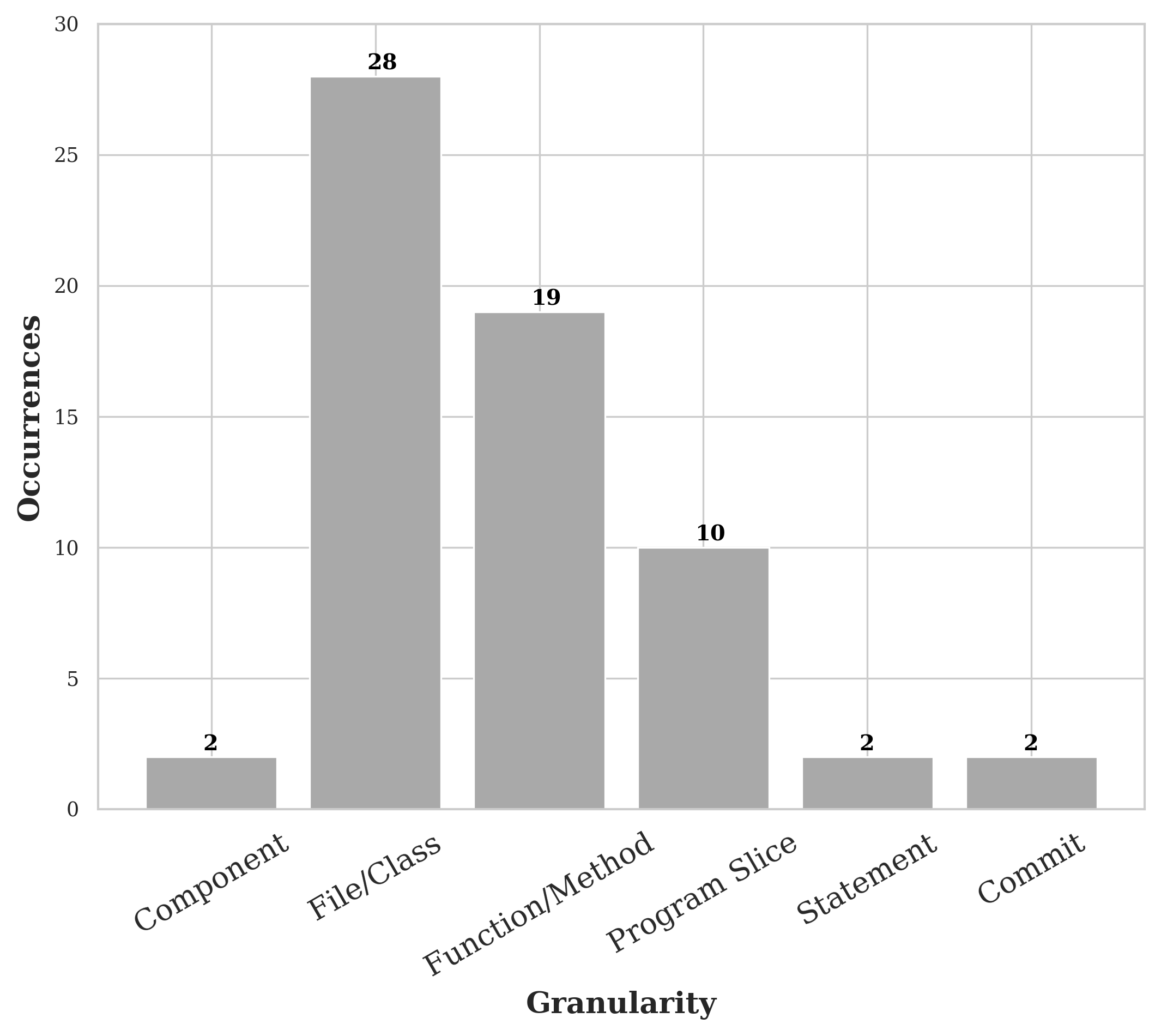}
  \caption{The number of primary studies for each level of granularity.}
  \label{fig:datasets_by_granularity}
\end{figure}

\textbf{Code module granularity.}
The granularity of vulnerability detection has a significant impact on a model and data collection. Depending on the granularity of the inputs used for an SVP model, it can either be used to direct testing efforts by predicting which large scale components are potentially at risk~\cite{morrison2015challenges}, or to explicitly detect fine-grained components that contain vulnerabilities~\cite{li2019survey}. In our set of primary studies, we identified six levels of granularity (in descending order of granularity): 1) component level, 2) file/class level, 3) function/method level, 4) program slice level, 5) statement level, 6) commit level. In Figure \ref{fig:datasets_by_granularity}, other than the component-level, the larger granularities are shown to be of more popularity. The file level has been considered as the standard granularity for SVP research [P12, P36, P46] and has been confirmed as actionable by developers [P5]. However, researchers have recently begun to favor finer granularities as they better enclose the scope of vulnerable code snippets, and are more easily inspected [P2, P3, P10, P15, P31, P32]. 

\textbf{Application contexts.}
There are three main vulnerability detection application contexts: within-project, cross-project, and mixed-project, in which each have different requirements. For within-project prediction, both the training data and the testing data come from the same project. In contrast, for cross-project prediction, there is an assumption that there is an insufficient amount of training data available in the target project. Therefore, labeled data from other source projects are used for training. More than one dataset can be collected to compensate for the inadequacy of labeled data in the target project. Mixed-project prediction is a special case of the cross-project setting. The labeled data from multiple projects are combined together to produce sufficient data for both model training and testing. This differs from the cross-project setting in which data from other projects only comprise the training dataset. 

Within-project prediction has proven to be the more popular prediction context, with 52\% (32 out of 61) of the primary studies forming their datasets from a singular project. This is because researchers have considered within-project prediction to be the standard use case of SVP [P5, P36]. Furthermore, cross-project prediction has often performed poorly due to differences in data distribution of the source and target project(s) [P1, P12, P26, P39, P56]. Only 18 of the primary studies considered cross-project prediction, 13 of which also considered within-project prediction. However, due to the various data issues that we will discuss in Section \ref{sec:data_challenges}, several researchers have considered the reuse of existing datasets from other projects to be a necessity [P1, P10, P18, P21, P22, P26, P31, P37, P41, P43, P56]. Twenty three of the primary studies (38\%) performed mixed-project prediction. 

\subsection{Data Collection}
SV datasets can be categorized into three main areas based on the type of data sources used to generate the code modules: real-world data, synthetic data, and mixed data. The type of data is the main influence on how the data is collected. 

\textit{\textbf{Real-world data.}} Both the code and the corresponding vulnerability annotations have been derived from real-world repositories. The code has been typically collected for projects hosted on repository hosting sites, such as GitHub~\cite{GitHub}, or through a different version control system. Experiments conducted using this data are usually considered to be better representative of industrial application because of the reflection of the complexities of the real-world vulnerabilities \cite{chakraborty2021deep}.

\textit{\textbf{Synthetic data.}} The vulnerable code examples and the labels have been artificially created. The examples from these data sources are synthesized using known vulnerable patterns. Synthetic datasets include SARD~\cite{SARD}, OWASP Benchmark~\cite{OwaspBenchmark}, and SQLI-Labs~\cite{SQLiLabs}. These datasets were originally used for evaluating traditional static and dynamic analysis based vulnerability prediction tools, due to their large test suite size and noise-free information. 

\textit{\textbf{Mixed data.}} Several researchers have opted to create datasets by merging both real-world and synthetic data sources [P32, P44, P48]. This is typically done to achieve a sufficient dataset size whilst maintaining a certain level of real-world representation. Mixed datasets have been primarily constructed for DL-based studies [P32, P44], which are particularly data hungry in comparison to ML. 

\begin{figure}[ht]
  \centering
  \includegraphics[width=0.9\linewidth]{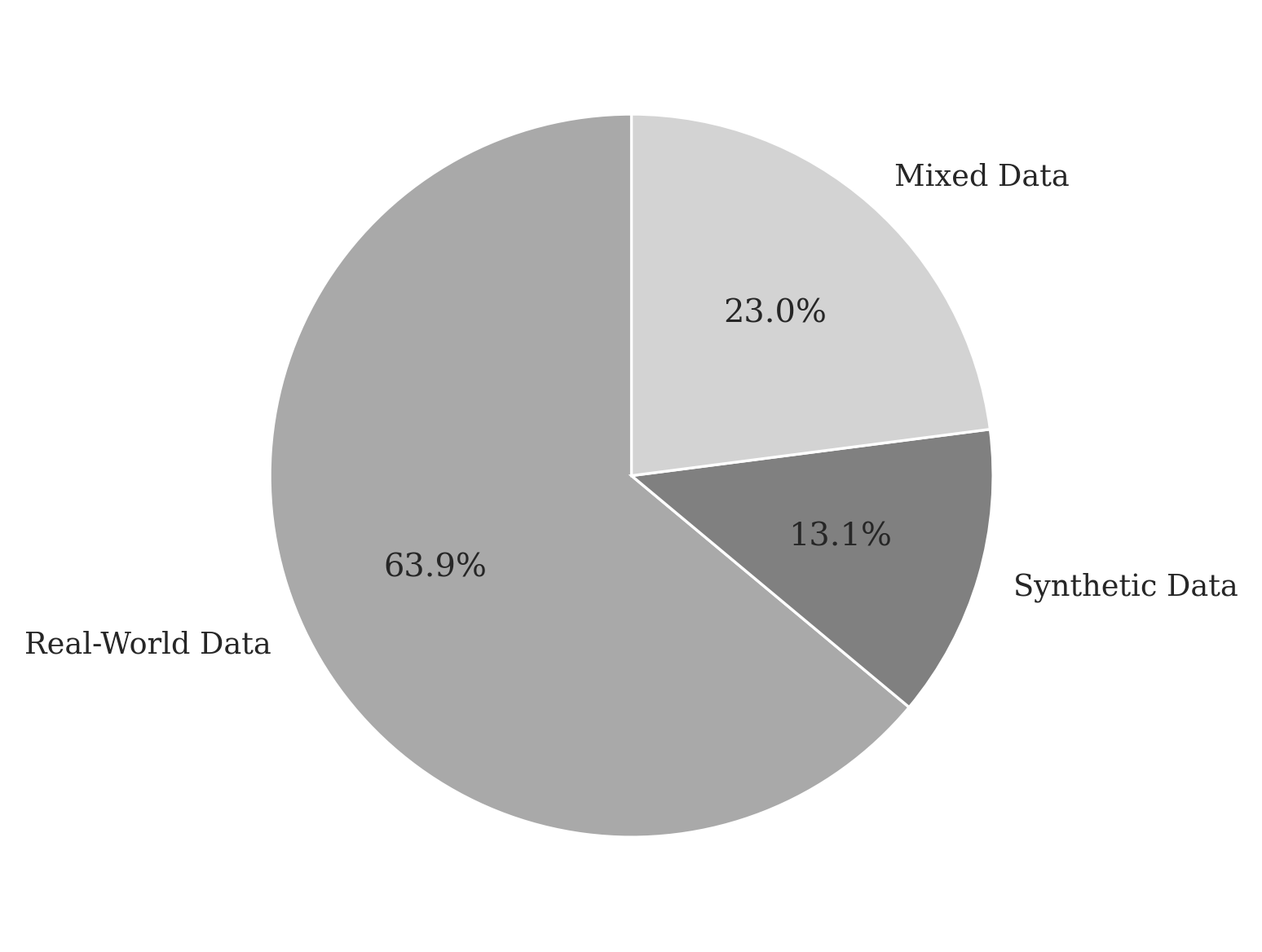}
  \caption{The proportion of primary studies for each data source type.}
  \label{fig:data_types}
\end{figure}

Figure \ref{fig:data_types} displays the proportion of primary studies that use each data type. Real-world datasets have been considered the \textit{de facto} type of data sources in this domain by researchers, primarily as they better represent real-world scenarios than synthetic examples~\cite{chakraborty2021deep}. The representativeness of a dataset is an important consideration as it helps improve the generalizability and validity of findings~\cite{ivarsson2011method}. Studies have further claimed that the code patterns of synthetic test cases follow a similar coding format, failing to reflect the characteristics of code patterns in production environments [P2 ,P3, P9, P29, P31, P39, P42, P43, P61].

However, there are two primary positive traits of synthetic datasets. Firstly, there are a much larger number of labeled synthetic samples that are able to be created in comparison to real-world examples [P8, P23, P34, P38, P39, P44, P54]. SVP is a data-hungry process that requires a sufficient amount of training data~\cite{hanif2021rise}. The existence of their labels also significantly reduces the effort of data preparation [P3, P6, P10, P39, P40]. Secondly, as the code samples are generated with their labels, the labels are cleaner and more reliable than data extracted from noisy real-world repositories [P39, P40], for reasons discussed in the following section. 

\subsection{Data Labeling}

\begin{figure}[ht]
  \centering
  \includegraphics[width=0.9\linewidth]{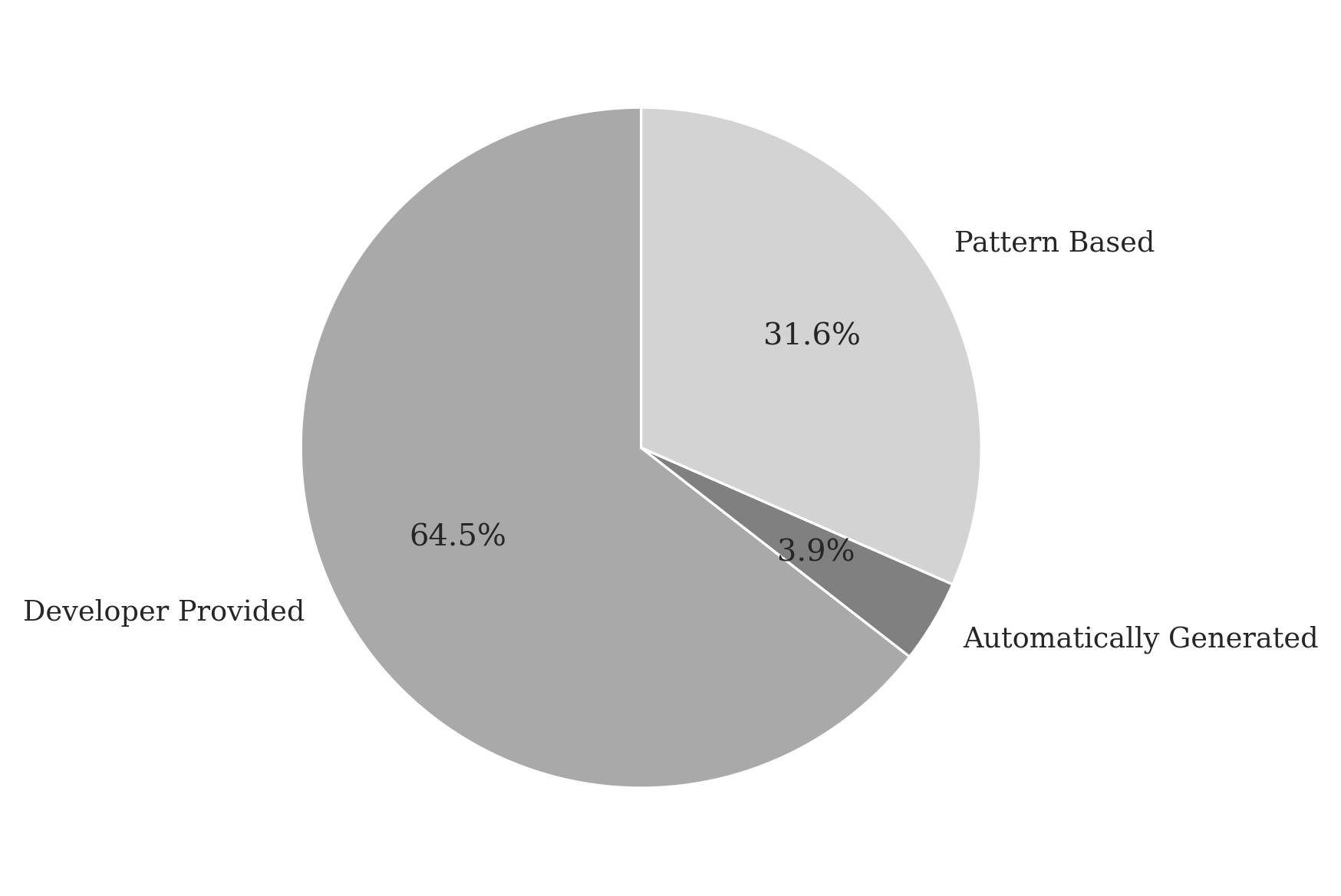}
  \caption{The ratio of SV data label sources. A study may use more than one label source.}
  \label{fig:label_types}
\end{figure}

For SV data, labels categorize whether a code module is vulnerable or not. Data labeling involves extracting data labels using an external source or tool, to assign to the collected code modules. We observed three SV label sources, that align with the findings of Chakraborty et al.~\cite{chakraborty2021deep}: developer-provided, automatically generated, and pattern-based. Figure \ref{fig:label_types} displays the ratio of these three SV label types. The choice of labeling approach is often dependent on the data source type collected, as outlined in Section \ref{sec:data_prep}.2. Hence, the relative frequency of each method is similar to that of Figure \ref{fig:data_types}. 

\begin{figure}[ht]
  \centering
  \includegraphics[width=0.9\linewidth]{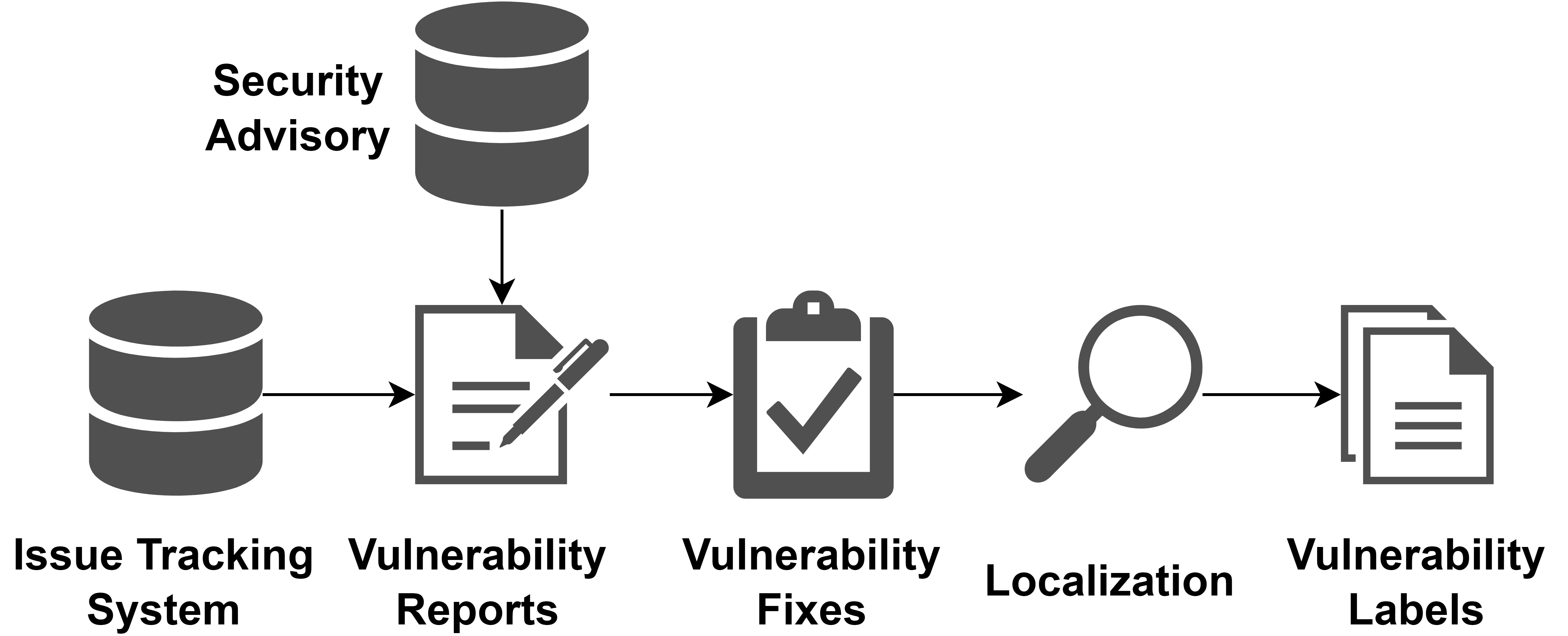}
  \caption{The SV labeling process using developer provided SV labels.}
  \label{fig:data_labeling}
\end{figure}

Figure \ref{fig:data_labeling} displays the labeling process using developer provided labels. These labels have been extracted from SVs that were identified by developers and reported in security advisories or issue tracking systems, such as NVD~\cite{NVD}, Jira~\cite{Jira} or Bugzilla~\cite{Bugzilla}. Whilst these information sources are usually accurate, it is not the same as developers hand-labeling code modules, as patches do not directly equate to labels. Researchers have expended significant effort to trace the label source to the code modules [P2, P18, P29, P43, P61]. This involves identifying the relevant vulnerability reports, extracting code fixes, and localizing these fixes to the relevant granularity and code modules, with each step introducing potential noise to the labels. Labeling of the non-vulnerable class is also quite subjective in this scenario, as there is no associated label source for this class. Consequently, it is a form of weak supervision~\cite{hernandez2016weak} that can introduce inadequacies into the data.

Automatically generated labels have not relied on a third-party label source, but instead have used an additional oracle to provide labels directly for the collected code modules. Two of the primary studies used static analysis tools to provide labels to code datasets [P37, P46]. This approach is noticeably the least considered as it heavily relies on the accuracy of the oracle used for labeling.

Pattern-based labels have been obtained for synthetic data sources, that use SV patterns to generate both the code modules and labels. The goal of an SVP model applied to this synthetic data, is to re-learn the patterns that generated the code modules of each class. These labels are largely considered noise-free, unlike the other two approaches, as the modules and labels are inherently connected. 

\subsection{Data Cleaning}
Data cleaning is the fourth step of data preparation. Whilst data cleaning is important, it is the only data preparation step that is non-essential. Collected code can be labeled and used immediately if it is extracted in an appropriate format. Hence, we observed that not all of the primary studies have discussed this step; nor has it been discussed in as much detail as the other steps. 

Code modules are the raw data for SVP, so the data cleaning step involves processing collected code modules so that they are in an appropriate format for feature extraction. Table \ref{tab:data_cleaning_approach} lists the common data cleaning approaches that we observed from the reviewed primary studies. The cleaning approaches fall under three issue types: irrelevant code, code noise, and duplication.

\begin{table}[ht]
  \caption{Common data cleaning approaches for SVP datasets.}
  \label{tab:data_cleaning_approach}
  \begin{tabular}{>{\raggedright\arraybackslash}p{2cm}>{\raggedright\arraybackslash}p{6cm}}
    \hline
    \textit{Issue} & \textit{Cleaning Approaches}\\
    \hline
    
    \multirow{3}{*}{\textbf{Irrelevant Code}} & \tabitem Remove code that is not of the target programming language(s).\\
    & \tabitem Remove code that is not of the target granularity.\\
    & \tabitem Remove irrelevant code files: test cases, third-party-code, scripts and make files.\\
    \hline
    
    \multirow{3}{*}{\textbf{Code Noise}} & \tabitem Remove blank lines, non-ASCII characters and comments from the code.\\
    & \tabitem Ignore code with syntax issues or errors.\\
    & \tabitem Replace the user-defined variable and function names with generic tokens.\\
    \hline
    
    \multirow{2}{*}{\textbf{Duplication}} & \tabitem Remove highly correlated items.\\
    & \tabitem Remove the duplicated code.\\
    \hline
    
\end{tabular}
\end{table}

\textbf{Irrelevant Code.} Firstly, irrelevant code has been commonly removed from the collected code modules. For all of the primary studies, this involves removing any code that was not of the target programming language, and removing any code that did not fall under the target granularity, such as code not contained in functions for the function-level granularity. Some studies also removed code that was not of relevance to the project or not at risk, such as test cases, third-party code modules, code scripts, and make files [P1, P4, P30, P56].

\textbf{Code Noise.} Secondly, some studies have further processed the collected code to remove the potential noise in the data that may impact the created features. These steps are only of value to the studies that extract the features directly from the code tokens, while software-metrics are usually robust to such code noise. Several studies removed comments, blank lines and non-ASCII characters [P1, P10, P20, P28, P32, P34, P44, P56], as these are not relevant to SVs. Some studies also replaced user specific tokens, such as user-defined variables, function names, and string literals, with generic tokens to increase the homogeneity of the features [P7, P23, P28, P29, P32, P50, P56]. Some studies also removed the code that they found to have syntax issues or errors [P9, P50], as this may later impact the feature extraction efforts. 

\textbf{Duplication.} Thirdly, several studies have attempted to remove duplicated code modules, as duplicate entries may introduce bias into a model~\cite{domingos2012few}. Duplicate code modules can be present due to multiple of the same code file, snippet, or software version being collected [P1, P8, P15].

\section{Data Challenges (RQ2)}
\label{sec:data_challenges}

This section identifies the data challenges that researchers have reported in the primary studies, in relation to the data preparation steps. As discussed in Section \ref{sec:data_synthesis}, we used thematic analysis to analyze the data quality issues that researchers have explicitly reported. These issues were coded, revised and merged by the first two authors of this study. The themes that we have identified are inspired by existing data quality dimensions~\cite{pipino2002data, sidi2012data}, as data challenges revolve around data quality. Table \ref{tab:data_challenges} details the key data challenges and considerations that we observed through our analysis. The data challenges can be summarized as pertaining to data Generalizability, Accessibility, labeling Effort, Scarcity, and both Label and Data Noise. 

\begin{table*}[t]
  \caption{Taxonomy of SV data challenges identified from the primary studies.}
  \label{tab:data_challenges}
  \resizebox{\textwidth}{!}{%
  \begin{tabular}{>{\raggedright\arraybackslash}l>{\raggedright\arraybackslash}l>{\raggedright\arraybackslash}p{6cm}>{\raggedright\arraybackslash}p{4cm}l}
    \hline
    \textbf{Theme} & \textbf{Challenge} & \textbf{Key Points} & \textbf{Paper ID} & \textbf{\#}\\
    \hline
    
    \multirow{7}{*}{\textbf{\textit{Generalizability}}} & \multirow{1}{*}{\textbf{Ch1:} Real-World Representation} & \tabitem Synthetic data is not representative of real-world code & P[2-3, 9, 29, 34, 39, 60] &  \multirow{7}{*}{43}\\
    \cline{2-4}
    & \multirow{3}{*}{\textbf{Ch2:} External Generalization} & \tabitem Data may be language specific & P[1-2, 11, 15, 20-23, 26, 30, 32, 38, 41-42, 44, 46-47, 55-57, 60-61] & \\
    & & \tabitem Data may be application or domain specific & P[1, 4-5, 11-13, 19-22, 30, 35-36, 40-42, 46, 52, 55-57, 59] & \\
    & & \tabitem Data may be specific to vulnerability type & P[15, 24, 26, 32, 38, 61] & \\
    \cline{2-4}
    & \multirow{3}{*}{\textbf{Ch3:} Completeness} & \tabitem SVs may span code modules & P[1, 3, 7, 10, 15, 21, 31, 45] & \\
    & & \tabitem Code representation may have limited scope & P[1, 3, 8, 13, 32, 44] & \\
    & & \tabitem SVs may be present in non-targeted code files & P[1, 12, 20, 46, 53, 57] & \\
    \hline
    
    \multirow{7}{*}{\textbf{\textit{Accessibility}}} & \multirow{1}{*}{\textbf{Ch4:} Cold-Start Problem} & \tabitem SVs are required to have originally occurred & P[4, 10, 21-22, 31, 37, 41, 56] & \multirow{7}{*}{25}\\
    \cline{2-4}
    & \multirow{2}{*}{\textbf{Ch5:} Data Entry Availability} & \tabitem Not all data entries are obtainable & P[1, 5, 11-13, 20-21, 30, 45, 49, 56] & \\
    & & \tabitem Usage of Version Control Systems is unstable & P[1, 12, 56] & \\
    \cline{2-4}
    & \multirow{2}{*}{\textbf{Ch6:} Data Privacy} & \tabitem Source code and SV data are required to be available & P[9-10, 16, 26-27, 32, 37-38, 45, 48, 61] & \\
    & & \tabitem Security advisories can be private or vague & P[1, 13, 56] & \\
    \hline
   
    \multirow{3}{*}{\textbf{\textit{Effort}}} & \multirow{1}{*}{\textbf{Ch7:} Labor Intensive} & \tabitem Manual labeling is highly time-consuming & P[2, 10, 18, 26, 29, 31, 39, 42-43, 61] & \multirow{3}{*}{14}\\
    \cline{2-4}
    & \multirow{2}{*}{\textbf{Ch8:} Expertise Requirements} & \tabitem Manual labeling requires high expertise & P[2, 18, 31, 43] & \\
    & & \tabitem Vulnerabilities are difficult to identify & P[20, 30, 33-34, 42-43] & \\
    \hline
   
    \multirow{2}{*}{\textbf{\textit{Data Scarcity}}} & \multirow{2}{*}{\textbf{Ch9:} Data Imbalance} & \tabitem Vulnerable samples are the extreme minority & P[1, 4-5, 8-9, 11-12, 14-15, 17, 19-20, 22, 24, 30-31, 33, 36, 40, 45, 50, 53-56, 60] & \multirow{5}{*}{32}\\
    \cline{2-4}
    & \multirow{1}{*}{\textbf{Ch10:} Number of Samples} & \tabitem Low number of vulnerability samples & P[1, 4, 10-11, 15, 20, 23, 29-31, 33-34, 36, 45-47, 50] & \\
    \hline

    \multirow{10}{*}{\textbf{\textit{Label Noise}}} & \multirow{2}{*}{\textbf{Ch11:} Incomplete Reporting} & \tabitem Latent, dormant or unresolved SVs can exist in the dataset & P[1, 4, 11-13, 16, 19, 21, 28, 30, 36-37, 47, 56, 59-61] & \multirow{10}{*}{31}\\
    & & \tabitem SVs can be silently patched & P[4-5, 12, 30, 33, 36, 40, 47, 52, 56] & \\
    \cline{2-4}
    & \multirow{4}{*}{\textbf{Ch12:} Localization Issues} & \tabitem Commit noise causes localization issues & P[1, 16, 29, 42] & \\
    & & \tabitem Data noise causes localization issues & P[2, 8, 13, 28, 32, 44] & \\
    & & \tabitem Bug reports do not document code location & P[28, 30, 33, 42, 56] & \\
    & & \tabitem Version tracking is complex and erroneous & P[1, 24, 30, 56] & \\
    \cline{2-4}
    & \multirow{4}{*}{\textbf{Ch13:} Erroneous labeling} & \tabitem Manual labeling can be inaccurate or subjective & P[11, 18, 40, 43, 52, 56] & \\
    & & \tabitem Static analysis tools label modules inaccurately & P[2, 46] & \\
    & & \tabitem SVs may not actually be exploitable & P[1, 52] & \\
    & & \tabitem Label quality is unknown & P[24, 28, 32] & \\
    \hline
    
    \multirow{7}{*}{\textbf{\textit{Data Noise}}} & \multirow{2}{*}{\textbf{Ch14:} Code Noise} & \tabitem Source code has stylistic differences or syntax issues & P[7, 9-10, 13, 18, 20, 23, 28-29, 32, 34, 38, 41, 50-51, 61] & \multirow{7}{*}{35}\\
    & & \tabitem Binary code is noisy & P[27, 48, 53] & \\
    \cline{2-4}
    & \multirow{3}{*}{\textbf{Ch15:} Redundancy} & \tabitem Some entries are indistinguishable between classes & P[49] & \\
    & & \tabitem Code versions and localization can add redundancy & P[1, 8, 12, 15, 19-20, 24, 30, 42, 46] & \\
    & & \tabitem Vulnerable samples have limited diversity & P[14, 28, 40] & \\
    \cline{2-4}
    & \multirow{2}{*}{\textbf{Ch16:} Heterogeneity} & \tabitem Data contains outliers & P[25, 42] & \\
    & & \tabitem Poor cross-project performance & P[1, 12, 26, 39, 56] & \\
    \hline
    
\end{tabular}%
}
\end{table*}

\subsection{Generalizability}
Generalizability describes the ability for data to extend to other contexts, both in terms of findings and application~\cite{sidi2012data}. Hence, this largely measures the external validity of the produced analysis, based on the data. This challenge primarily involves the data requirements step, as this is the phase where researchers determine the nature of their dataset. 

\textit{\textbf{Ch1: Real-World Representation.}} Most of the challenges described in Table \ref{tab:data_challenges} arise when using real-world data. Consequently, several researchers have opted to use synthetically created data to construct their models, as described in Section \ref{sec:data_prep}.2. Synthetic datasets are artificially crafted to address data challenges present in the real-world data; these data sources are accessible, large, low-effort, and less noisy. As such, they are an attractive option for researchers.

Given that synthetic data may not represent the real-world data, it is considered a big limitation that may render such a dataset unusable unless this limitation is addressed. Synthetic vulnerability examples are considered to be simpler, isolated, less diverse and cleaner than real-world vulnerabilities [P2, P3, P9, P29, P34, P39, P60]. Zheng et al. [P29] found that the use of synthetic data sources may significantly inflate the reported model performance in comparison to the models using real-world code. Hence, a model trained using synthetic data is unlikely to be able to detect complex real-world vulnerabilities, which require much deeper semantic understanding and reasoning~\cite{chakraborty2021deep}. Thus, real-world data is the more commonly used data source, as seen in Figure \ref{fig:data_types}.

\textit{\textbf{Ch2: External Generalization.}} Nearly all studies face external threats to validity of their findings inferred from a specific dataset. In terms of SVP research, this relates to the limited application scope of the selected study datasets. Datasets may be specific to, and hence have troubles generalising outside of: programming language [P1, P11, P15, P20, P21, P22, P23, P25, P26, P30, P32, P38, P41, P42, P44, P46, P47, P55, P56, P57, P60, P61], application or domain type [P1, P4, P5, P11, P12, P13, P19, P20, P21, P22, P30, P35, P36, P40, P41, P42, P46, P52, P55, P56, P57, P59], and SV type [P15, P23, P26, P32, P38, P61].

\textit{\textbf{Ch3: Completeness.}} Completeness is achieved when a dataset has all the relevant parts of an entity's information, which is sufficient to represent every meaningful state of a real-world system~\cite{sidi2012data}. However, the selected data for analysis can have a limited scope of the overall system, which makes their application context limited. This data-oriented consideration serves as a challenge for SVP, as it prevents these models from forming a ``complete'' solution. Firstly, if the selected granularity of code modules is too fine, researchers are forced to ignore vulnerabilities which span multiple modules [P1, P3, P7, P10, P31, P45]. For instance, function-level prediction is unable to predict more complex SVs that span multiple functions. The selected semantic representation of data may also not consider all sources of weaknesses in a software system [P3, P8, P32, P44]. For instance, Tian et al. [P3] and Li et al. [P8] only consider code snippets of library and API function calls, which would not cover all potential SVs in a system. Static source code is also unable to capture certain necessary dynamic code traits [P1, P13], such as crashes and memory leaks. Similarly, vulnerabilities may be present in code modules which are not of the target programming language of analysis [P1, P12, P20, P46, P53, P57]. In the modern development landscape projects commonly utilize multiple programming languages~\cite{kochhar2016}, but SVP models are largely targeted towards a single programming language of choice [P53].

\subsection{Accessibility}
Accessibility describes the ability to retrieve or obtain data from the target data sources~\cite{sidi2012data}. Challenges arise from difficulties in accessing the data, either of the raw code modules during the data collection step or of the data labels during the data labeling step. 

\textit{\textbf{Ch4: Cold-Start Problem.}} The cold-start problem is an issue originating from recommender systems in which a system is unable to draw inferences about incoming modules for which it has not yet gathered sufficient information~\cite{lika2014facing}. In terms of SVP, the cold-start problem has been particularly present, as to make future predictions, we require vulnerabilities to have originally occurred and to have been documented [P4, P10, P31]. This makes SVP largely infeasible for new or immature organisations [P10, P21, P22, P31, P37, P41, P56]. Furthermore, the acquisition of initial high-quality training data is a major issue, as seen in the other challenges. 

\textit{\textbf{Ch5: Data Entry Availability.}} Since a majority of SV data have been obtained through mining open source repositories, not every part of a system would be necessarily accessible to researchers. For instance, some source code might have been unavailable due to lack of inclusion in public repositories [P1, P11, P12, P20], or unobtainable due to other unstated technical reasons [P5, P21, P45]. Similarly, some vulnerability reports might not have been able to be localized to code modules due to issues in the automatic or manual localization methods [P1, P13, P30, P49, P56]. 

Some researchers have even pointed out that the reliance on a version control system to track code modules causes issues in itself, as consistent usage of a version control system is unstable [P1, P12, P56]. Version control systems were only widely adopted in 2005 with the introduction of \textit{git}, hence data before this date would be irretrievable [P12, P56]. Furthermore, organisations might switch the version control system they were using, losing previous software history [P1]. 

\textit{\textbf{Ch6: Data Privacy.}} The potential commercial sensitivity of both software code and SV reports means that organizations are often not willing to share private-source code or data to researchers\cite{bosu2013taxonomy}. This data privacy creates many data accessibility issues. Firstly, several researchers have observed that commercial systems have not provided their source code [P9, P10, P26, P32, P45, P48, P61], making SVP via source code on these systems infeasible. Furthermore, organisations and practitioners might desire to limit the availability of their security advisories by making them private or vague [P1, P13, P56]. By concealing information about vulnerabilities, it is theoretically harder for an attacker to exploit a system. Similarly, an organisation might not even maintain a public security advisory or document SVs [P16, P27, P37, P38]. 

To represent real-world data, researchers have often surreptitiously avoided this issue through the use of open-source repositories that have public vulnerability records. Without open sources, data retrieval and reporting can become very difficult due to commercial sensitivity. Whilst this is valid, open-source data is not representative of software engineering practices as a whole; it is unknown whether the derived observations will generalize to private-source code and practices. Only two out of the 61 primary studies used private source data [P5, P11], both of which suffered from data entry availability (\textbf{Ch5}) as a result.

\subsection{Effort}
Effort describes the amount of human-effort required to label a dataset~\cite{sidi2012data}. The standard approach for traditional supervised learning has been to have a subject matter expert hand-label a dataset. However, this is largely infeasible for SV data due to the extreme effort requirements. As such, many researchers have skirted around this challenge by using synthetic labeled data sources or reusing existing datasets. Consequently, this theme was actually the least mentioned by primary studies, as researchers that reused datasets often did not report or discuss the effort. 

\textit{\textbf{Ch7: Labor Intensive.}} Labeling code or bug reports as SV-related is a non-trivial task, which makes it highly labor intensive when coupled with the sheer number of modules to examine [P2, P18, P29, P43, P61]. Dowd et al.~\cite{dowd2006art} estimated that one hour of security review can cover an average of 500 lines of code. However, most modern software systems contain millions of lines of code, which makes the required man-hours infeasible. Zhou et al. [P2] stated that it took 600 man-hours to manually curate their SV dataset. Manual labeling is ultimately the most reliable labeling approach however, due to the large amount of noise for automated labeling (\textbf{Ch13}). 

\textit{\textbf{Ch8: Expertise Requirements.}} Secure code review requires significant security expertise [P2, P18, P31, P43]. To successfully perform secure code review, a practitioner/researcher must have the capability to memorize and recognize thousands of security-related patterns and concepts~\cite{barnum2005}, and this list of required knowledge is continually growing. Furthermore, it has been highly difficult to identify SVs in comparison to regular defects [P20 , P30, P33, P34, P42, P43], as they do not necessarily represent functional bugs and are thus hard to verify.

\subsection{Data Scarcity}
Data scarcity refers to the extent to which the quantity or volume of the available data is appropriate for the task at hand~\cite{wang1996beyond, pipino2002data}. As SVP is a data-driven process, it requires large volumes of data~\cite{domingos2012few}. For SV data, this challenge largely represents the low number of real-world vulnerable examples available. Whilst codebases are often sufficiently large, the number of identifiable vulnerable modules in a codebase is relatively small~\cite{zimmermann2010searching}. 

\begin{figure}[h]
  \centering
  \includegraphics[width=\linewidth]{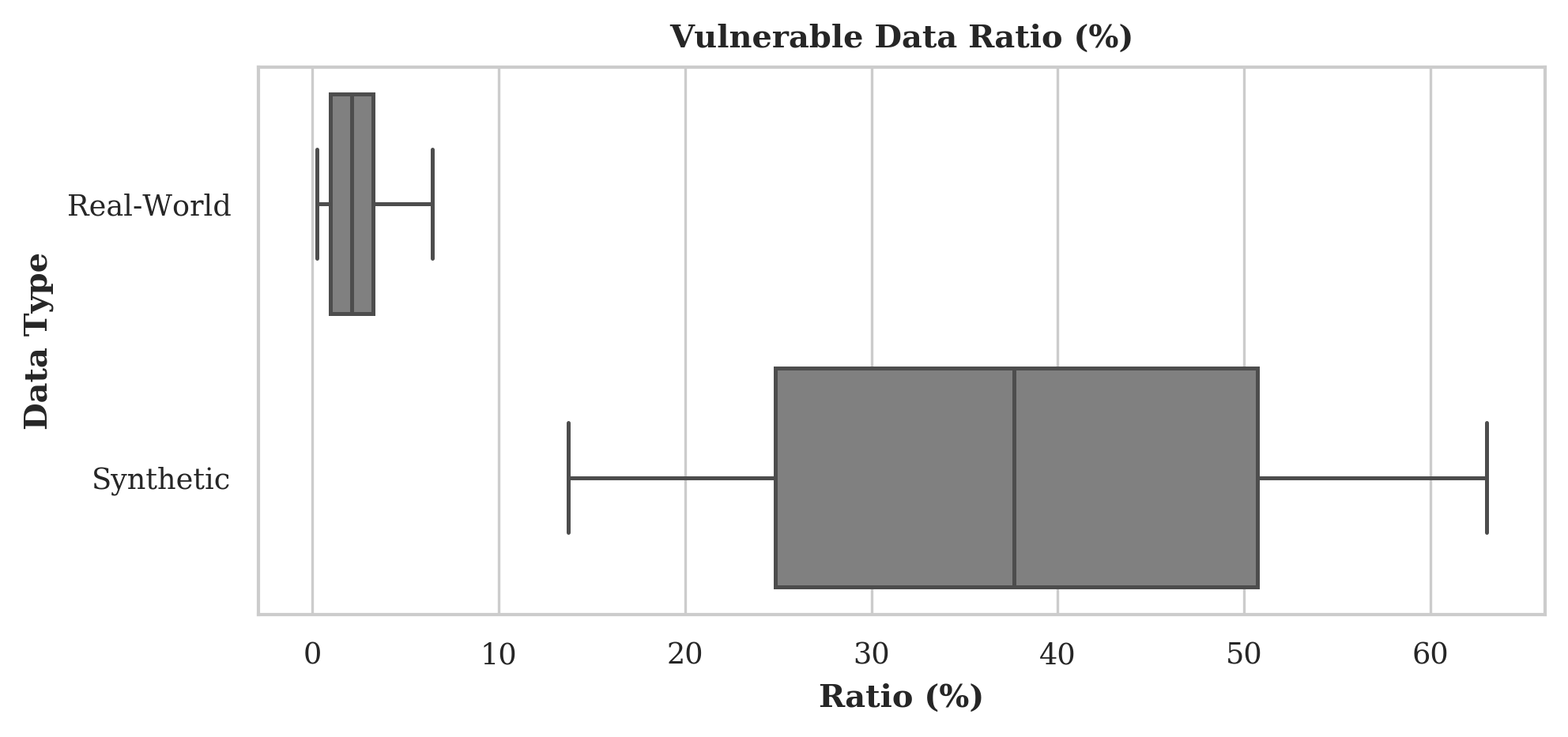}
  \caption{The percentages of vulnerable files in datatasets utilized in the primary studies.}
  \label{fig:data_proportion}
\end{figure}

\textit{\textbf{Ch9: Data Imbalance.}} The severe imbalance of vulnerable modules to non-vulnerable modules has been a challenge reported by many researchers using real-world datasets. To quantify this issue, we report the percentages of vulnerable modules in each dataset utilized in the primary studies, displayed in Figure \ref{fig:data_proportion}. We display the real-world and synthetic datasets separately, as the synthetic datasets have been artificially constructed to over-represent the vulnerable class. Real-world datasets which were artificially altered to be balanced, or merged datasets consisting of both synthetic and real-world examples are excluded from Figure \ref{fig:data_proportion}. 

This has been a considerable issue for SVP research, as learning-based models are optimized to perform on balanced classes~\cite{he2009learning}. The severe class imbalance issue present in real-world SV data can lead to biased classifiers that favor the non-vulnerable class [P15, P19, P24, P30, P31, P33, P50, P54]. This challenge has been referred to as \textit{finding a needle in a haystack}~\cite{zimmermann2010searching} and is notably unique to security defects. Shin and Williams [P33] observed that the number of reported faults was seven times larger than that of vulnerabilities. Nguyen and Tran [P24] observed that class imbalance increased as the module granularity became more fine-grained. 

\textit{\textbf{Ch10: Number of Samples.}} Similar to \textbf{Ch9}, the severe imbalance of data leads to a very low overall number of samples for the vulnerable class. This has been a major blockade for SVP research as learning-based methods have strict data requirements; they need a large quantity of historical cases to learn from. It is common knowledge in the ML community that \textit{more data beats a cleverer algorithm}~\cite{domingos2012few}. Several studies have particularly exacerbated this issue for DL-based methods, which require even greater data size [P15, P23, P29, P34, P45, P47, P50]. This is an emerging challenge given the rise of DL-based methods for SVP~\cite{lin2020software}.

\subsection{Label Noise}
Label noise relates to whether the labels in a dataset accurately represent the ground truth and are free of error~\cite{wang1996beyond, pipino2002data, bosu2013taxonomy}. As SV datasets have been rarely hand-labeled by subject matter experts (excluding synthetic datasets), but instead mined from historical artefacts, a large amount of noise has been introduced into the SV labels. This severely impacts the reliability of SV data. 

\textit{\textbf{Ch11: Incomplete Reporting.}} A major problem of using historically reported vulnerabilities to label real world data is that we only have a label source for the vulnerable class, and simply treated the remaining labels as non-vulnerable, creating uncertainty in the labels for the non-vulnerable class. Modules in the non-vulnerable class are not actually confirmed to be clean, just that no vulnerabilities have been historically reported. However, in reality, there can be dormant, latent or unreported vulnerabilities existing in these modules [P1, P11, P12, P13, P16, P19, p21, P28, P30, P36, P37, P56, P59, P60, P61]. Hence, the non-vulnerable class may be better considered as unlabeled [P21, P47]. 

Similarly, researchers are limited to the use of \textit{fixed} vulnerabilities as a label source [P52]. Unresolved vulnerabilities have been rarely disclosed as they can be exploited by attackers. Hence, during data labeling, these unresolved SVs would actually be contained in the non-vulnerable class. 

Furthermore, the reliance on vulnerabilities to be reported has created a temporal issue for data collection. SVs usually take time to be detected and fixed, and hence a different number of SVs will be documented depending on the time of data collection [P21, P36]. Jimenez et al. [P36] observed that the performance of models significantly decreased when only using vulnerabilities observed before the time of model training. 

Using bug reports for labeling also has created a reliance on developers or organisations to have thorough reporting practices. However, in reality, some organisations have patched some vulnerabilities ``silently'' [P36, P40, P56], without any documentation provided for bug reports or security advisories. Furthermore, the difference between SVs and non-security related faults can sometimes be minimal [P5, P12, P30, P33]. Hence, many security defects have not been reported by developers as such [P4, P52]. An organisation may also use multiple bug reporting systems; a reliance on just one data label source (i.e., NVD) would be incomplete [P36, P40]. 

\textit{\textbf{Ch12: Localization Issues.}} Bug reports and SV records have not always documented the location of SVs, posing a large challenge for the retrieval of SV data [P56]; the vulnerable code modules are often not explicitly listed. Hence, researchers have often relied on the use of patches to trace the location of a vulnerability. This is flawed as not every documented vulnerability has an associated patch [P28, P30, P33, P42, P56], as it may be concealed for privacy or not yet resolved. Furthermore, patches may not always properly disclose the true location of an SV [P1], as patches may instead provide workarounds for separate modules, rather than a fix of the underlying problem. Jimenez et al. [P12] found that only 75\% of vulnerability reports had an associated fix, which leaves the remaining 25\% of reports untraceable. Vulnerability reports are often incomplete and missing references~\cite{anwar2020cleaning}. 

Furthermore, tracking SV location from a bug fix has been non-trivial. Version information in bug reports is often unreliable [P1, P24, P30, P56], as identifying affected software releases for a vulnerability is highly difficult. Due to the evolving nature of code, the assumption that all previous versions of code were also vulnerable is invalid [P24], and reporting vulnerabilities in prior versions is inaccurate as developers have little benefit from  expending efforts to do so [P1]. Furthermore, commits can be noisy [P1, P16, P29, P42] due to tangled code changes. A vulnerability fixing commit may not exclusively patch a vulnerability \cite{herzig2016impact}; other functional changes may be included. A vulnerability fix can also be buried as part of a larger commit including non-security changes. Herzig et al.~\cite{herzig2016impact} showed that tangled commits had significant impacts on defect labeling, with an average of 16\% of files being mislabeled as a result.  

\textit{\textbf{Ch13: Erroneous labeling.}} Label inaccuracies can additionally come from various other sources. Firstly, with noisy labels stemming from \textbf{Ch11}, as well as errors arising from localizing labels to code modules (\textbf{Ch12}), manual labeling has been a common approach to help ensure data quality. However, this approach is erroneous in itself. As discussed in \textbf{Ch8}, manual labeling is difficult and requires high expertise. Hence, it is inevitably an error-prone task [P18, P40, P43, P52]. Furthermore, the process of labeling modules as vulnerable or not can even be quite subjective [P11, P56]. There is no clear definition of the difference between a vulnerability and a fault, and hence this distinction can be nebulous to a human [P30]. For instance, if a regular function calls a vulnerable function, it is unclear whether this function should also be considered vulnerable [P1]. 

Some studies have utilized static analysis tools and methods to achieve automatic data labeling without the need for historical vulnerability reports [P12, P37, P46]. However, researchers have observed these methods to be highly inaccurate and hence introduce considerable noise into data labels [P2, P46]. This process can also be flawed from a motivational perspective, as the SVP model is simply relearning the patterns used by the static analysis tool to infer the labels. 

Patched vulnerabilities may not actually be exploitable in the real-world either [P1, P52]. Developers may incorrectly hypothesize security weaknesses, or err on the side of caution. This adds unreliability to the accuracy of the labels in the data source, as the SV labels may actually be benign.  

Another large and open challenge has been the lack of measures to quantify label quality [P24, P28, P32]. There is no trivial way to measure or quantify the aforementioned label noise issues. Furthermore, with the reliance on historical artefacts for labeling, some sources of label noise are undetectable or unverifiable, e.g., SVs remain dormant (\textbf{Ch11}) until they have been detected. This severely impacts the reliability of SVP research, as it is unclear whether the findings have been made using valid data.

\subsection{Data Noise}
Finally, data noise refers to the noise within the raw data entries~\cite{sidi2012data}; code modules for the purposes of SVP. Noise and inaccuracies in these modules may negatively affect the data and any produced features used to train a model, lowering the potential efficacy of that model.

\textit{\textbf{Ch14: Code Noise.}} SVP uses code as the raw data source. However, source code is noisy, which consequently impacts the effectiveness of any produced model. Developers have different coding styles and naming conventions [P28, P34, P38], which adds inconsistency. Li et al. [P61] further identified that different projects may have different code quality due to differences in coding practices and guidelines. Real-world code can also contain syntax issues [P9, P50]. These sources of noise can severely impact the versatility of produced SVP models, as they may instead learn specificities of particular coding styles and syntax. 

Furthermore, binary code is usually much noisier than regular source code. Binary code snippets can be difficult to trace and identify [P27, P53], or interspersed code and variables can become indistinguishable [P27, P48]. 

\textit{\textbf{Ch15: Redundancy.}} Redundancy refers to undesirable duplication in a dataset. Too much redundancy in the training data, can lead to bias and overfitting for an SVP model. The major source of redundancy in SV datasets has come from code modules having several different versions and revisions. Datasets that consider versions separately can introduce redundancy into the code entries, as the majority of the code usually stays constant between revisions [P20]. Vulnerable labels can also be duplicated over several versions, as modules can remain vulnerable for an extended period of time [P12, P19, P24, P30, P46]. Code branches are another potential source of redundancy to labels and modules as the majority of code and data is often duplicated across branches [P1, P42]. Automatic extraction of code snippets and program slices can additionally introduce duplication [P8], as duplicate program slices can be created from different entry points. 

Vulnerable samples can also not be very distinct from each other [P14, P40], due to consistent SV type or exploit patterns, which limits the learning capacity of an SVP model. This is particularly present for synthetically created samples [P28]. Another issue is that vulnerable and non-vulnerable entries have limited diversity [P49]. Vulnerability patches can only alter a few lines of code, making the majority of the module consistent between its vulnerable and non-vulnerable versions. This is a particularly significant issue as if the model cannot learn these subtle distinctions, it will produce high false positive/negative rates. 

\textit{\textbf{Ch16: Heterogeneity.}} Code modules have been observed to be highly heterogeneous, which negatively impacts the diverse application of the produced SVP modules. For example, the coding conventions of one code module often do not match another, due to differences in authorship, functionality or coding style. Learning-based methods operate best when the data, especially the training and test distributions, are homogeneous so that the learnt patterns can be applied uniformly~\cite{domingos2012few}. However, researchers have observed data distributions to be irregular or containing outliers, hence requiring normalization to reduce irregularities [P25, P42]. Similarly, several researchers have observed that SVP models perform poorly in a cross-project setting [P1, P12, P26, P39, P56], due to the heterogeneity of these datasets; the coding conventions and functionality of one project rarely mirror another.

\section{Data Challenge Solutions (RQ3)}
\label{sec:data_solutions}

\begin{figure}[h]
  \centering
  \includegraphics[width=\linewidth]{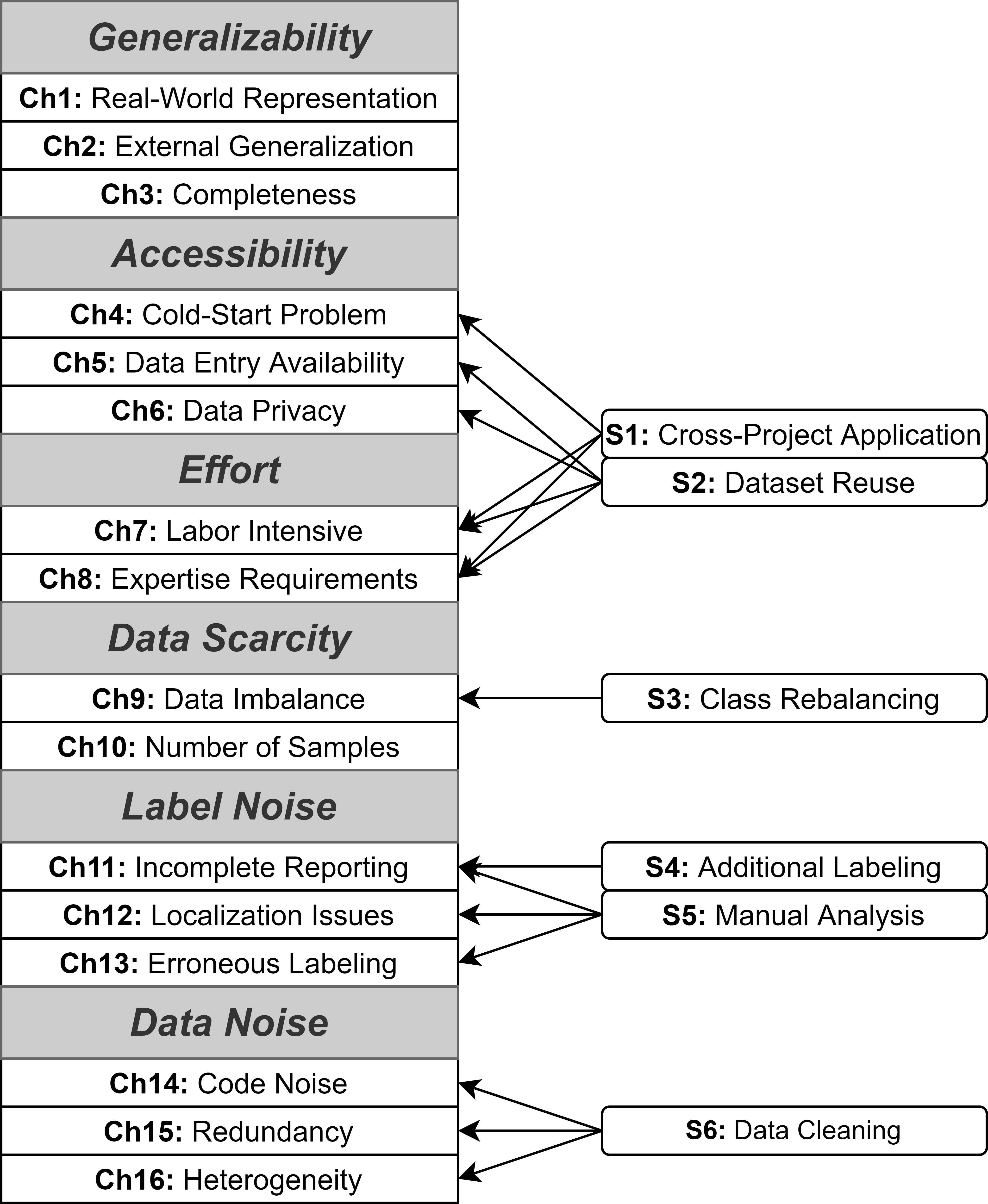}
  \caption{A mapping of solutions onto challenges.}
  \label{fig:data_solutions}
\end{figure}

In this section, we present the various solutions that researchers have presented in the reviewed studies to solve data challenges or help improve data quality. Figure \ref{fig:data_solutions} displays the main areas of the solutions that we have identified. We again used thematic analysis, described in Section \ref{sec:data_synthesis}, to identify the categories of the identified solutions. We mapped the solutions to the data challenges based on the data challenge themes that the solutions were connected to when reported in the primary studies. 

We provide an overview of the main solutions that have been considered for the identified data preparation challenges. As discussed in \textbf{Ch1}, the majority of the data challenges arise from the use of the real-world data; challenges for Accessibility, Effort, Data Scarcity and Label Noise are largely unique to the real-world datasets. Consequently, the majority of the data solutions are similarly aligned with the real-world data challenges. We note that not every primary study provided remediation for the reported data issues. 

We also note that several of the data solutions were provided to these challenges at a model-level, rather than from the data perspective. For instance, several studies supported the use of ensemble models as they are more robust and hence less susceptible to noise and class imbalance [P42]. Furthermore, feature processing techniques, such as feature selection, were another method used to help remove data noise and increase generalizability [P22, P25, P40, P42, P47, P55]. However, we focus on the data preparation steps, so an analysis of these model-level solutions is out of the scope of this SLR. 

\textit{\textbf{S1: Cross-Project Application.}} As discussed in \textbf{Ch4}, researchers have been motivated to overcome the cold-start problem and cost of acquiring a dataset. This has led to several efforts to apply cross-project SVP models for which a previous project's dataset is used to train an SVP model that can be applied to a new project. Due to the extreme cost of acquiring data (\textbf{Ch7\&8}), the cold-start problem is thought to be further exacerbated for SVP. Hence many researchers considered cross-project SVP as an essential solution to this issue [P1, P10, P18, P21, P22, P26, P31, P37, P41, P43, P56]. 

Whilst this solution does solve the challenges that it seeks to address, it also adds challenges by introducing heterogeneity to the data (\textbf{Ch16}), which consequently impacts a model's performance. Furthermore, underlying problems in the original data source can still be present.

\textit{\textbf{S2: Data Reuse.}} With the significant challenges in the data labeling effort (\textbf{Ch7\&8}), coupled with the challenges in data accessibility (\textbf{Ch5\&6}), it is a more efficient choice to reuse or augment the existing datasets. This not only significantly saves time and effort in data preparation, but also enables researchers to conveniently evaluate and benchmark performance on the same datasets.  As dataset reuse greatly reduces effort, it is more desirable than self-construction of a dataset. Hence, researchers have often reused the existing datasets when available; 28 out of the 61 reviewed studies reused or augmented the existing datasets in some manner.

Like \textbf{S1}, whilst this solution certainly solves the challenges regarding Accessibility and Effort, it does not address the problems in the original dataset like Generalizability, Data Scarcity, Label Noise and Data Noise. However, many researchers have seemingly assumed the validity of the prior datasets, as the studies that utilized former datasets often did not discuss other data challenges.

Data reuse is achieved through data sharing efforts of other researchers. Several researchers made their datasets publicly available for use to assist in construction of SVP models by researchers and practitioners [P1, P8, P10, P14, P16, P26, P31, P32, P36, P38, P44, P52, P56, P61]. 

However, the actual usability of these datasets can create further issues for this solution. Firstly, the quality and reliability of the information provided in the existing datasets is unknown and unverifiable [P4, P8, P12, P20, P28, P38, P39, P40, P44, P54]. Researchers often prefer to use a dataset about which they have complete knowledge. Researchers also build their SVP models to fit a variety of application contexts, such as specific granularities, programming languages, or SV types. Hence, several researchers found the information provided in the existing datasets insufficient to be applicable to their desired applications' contexts [P2, P4, P12, P16, P23, P27, P31, P39, P42, P43, P48, P56]. 

Furthermore, although many SV datasets have been created, they are not necessarily available. Researchers have reported that the existing datasets are private [P2, P16, P31] or unavailable [P2, P12, P42]. We observed that five of the shared datasets from the reviewed studies have since become unavailable due to dead links. Similarly, researchers may not share the code used to produce a dataset, which impacts the verifiability and reproducibility. Such problems of availability or usability of the desired datasets can lead researchers to use self-constructed datasets. 

\textit{\textbf{S3: Class Rebalancing.}} The severe imbalance of vulnerable to non-vulnerable modules (\textbf{Ch9}) is reported by almost half of the reviewed studies (27 out of 61) as a significant data challenge. This imbalance leads to models that bias towards the majority class~\cite{domingos2012few}, and do not fairly consider the minority vulnerable class. As such, many studies have employed some form of class rebalancing to help remediate this issue. These rebalancing techniques fall into two main categories: Undersampling [P1, P8, P14, P15, P19, P20, P22, P30, P33, P52, P53 P55, P56, P60] and Oversampling [P8, P14, P17, P20, P36, P45, P56]. Undersampling is the process of removing samples from the majority class to match the size of the minority class, whereas oversampling duplicates or synthetically adds samples to the minority class until the size matches the majority class. Undersampling was the more popular technique used in the reviewed studies; 14 studies used undersampling in comparison to 7 that used oversampling. Researchers considered undersampling to be more standard [P22, P55], effective [P30, P60] or efficient [P19, P30, P60], in comparison to oversampling. However, we note that undersampling may not always be desirable as it reduces the overall amount of data, which can have significant impacts for SV data due to the low number of samples in the minority class (\textbf{Ch10}). Oversampling does not suffer from information loss, but adds redundancy to training data (\textbf{Ch15}), which can lead to overfitting. 

This solution is largely considered a necessity, as the positive impacts of class rebalancing have been widely agreed upon by the community. Furthermore, several studies have explicitly demonstrated the performance increase of models when trained on balanced data in comparison to imbalanced data [P7, P54, P60]. 

Alternatively, some researchers artificially constructed both their training and test datasets to be balanced [P61]. Balancing the test set like this artificially inflates the model performance, however, as the incoming real-world data will not be balanced [P1, P15]. This is a significant weakness of synthetic datasets that is expected to impact their real-world generalization, as they are constructed to over-represent the vulnerable class. 

\textit{\textbf{S4: Manual Analysis.}} The poor labeling provided from bug reports can add a lot of label noise and inaccuracies in the data. To combat this, researchers have often assisted the labeling process through manual analysis. Over 37\% of the reviewed studies (23 out of 61) assisted their data collection process with manual inspection. This allowed researchers to manually spot sources of label noise [P1, P2, P4, P7, P8, P10, P11, P14, P25, P26, P28, P29, P30, P31, P32, P42, P44, P35, P46], or made localization to code modules more accurate [P1, P2, P4, P7, P10, P14, P26, P31, P44, P45, P52]. It also helped researchers to obtain an understanding of the quality of their dataset [P13, P16, P32, P42]. 

Whilst manual labeling assistance can greatly improve the quality of a dataset, manual analysis is problematic in itself. It is highly effort intensive (\textbf{Ch7}), difficult (\textbf{Ch8}) and error-prone (\textbf{Ch13}). 

\textit{\textbf{S5: Additional labeling.}} Manual analysis can only resolve issues in the vulnerable class, as the use of bug reports only provides a source of labels for this class. In reality, however, dormant or latent vulnerabilities can introduce unverifiable inconsistencies in the non-vulnerable class (\textbf{Ch11}). Hence, several researchers have used automated methods to obtain additional labels. One such approach was to use static analysis tools to label modules, in an attempt to uncover latent vulnerabilities [P37, P46]. However, this process added considerable noise to the labels in itself [P2, P46]. Li et al. [P61] used a rule-based approach to obtain the non-vulnerable modules, by filtering out functions which may have security relevance. This decreased the chance of having latent vulnerabilities in the non-vulnerable set. 
As these additional labeling methods often introduce their own assumptions or inaccuracies, their effectiveness is unclear. 

\textit{\textbf{S6: Data Cleaning.}} Data noise problems can be addressed through data cleaning of code modules. Table \ref{tab:data_cleaning_approach} from Section \ref{sec:data_prep} lists common data cleaning techniques for SV modules. Although the real world code is usually much noisier than the synthetic code due to inconsistent coding styles [P28, P34, P38], cleaning can still be necessary for the synthetic data sources, depending on the method used to construct the vulnerable examples. Fang et al. [P50] found a large proportion of coding errors in the synthetic SARD and SQLI-Labs datasets, that they manually remediated. 

Data cleaning is one of the four data preparation steps \cite{amershi2019software}, and consequently this solution has been largely considered a necessity. For example, Zheng et al. [P28] showed that replacing user defined strings with generic tokens improves a model's performance.

\section{Recommendations}
\label{sec:recommend}

\begin{figure*}[h]
  \centering
  \includegraphics[width=\linewidth]{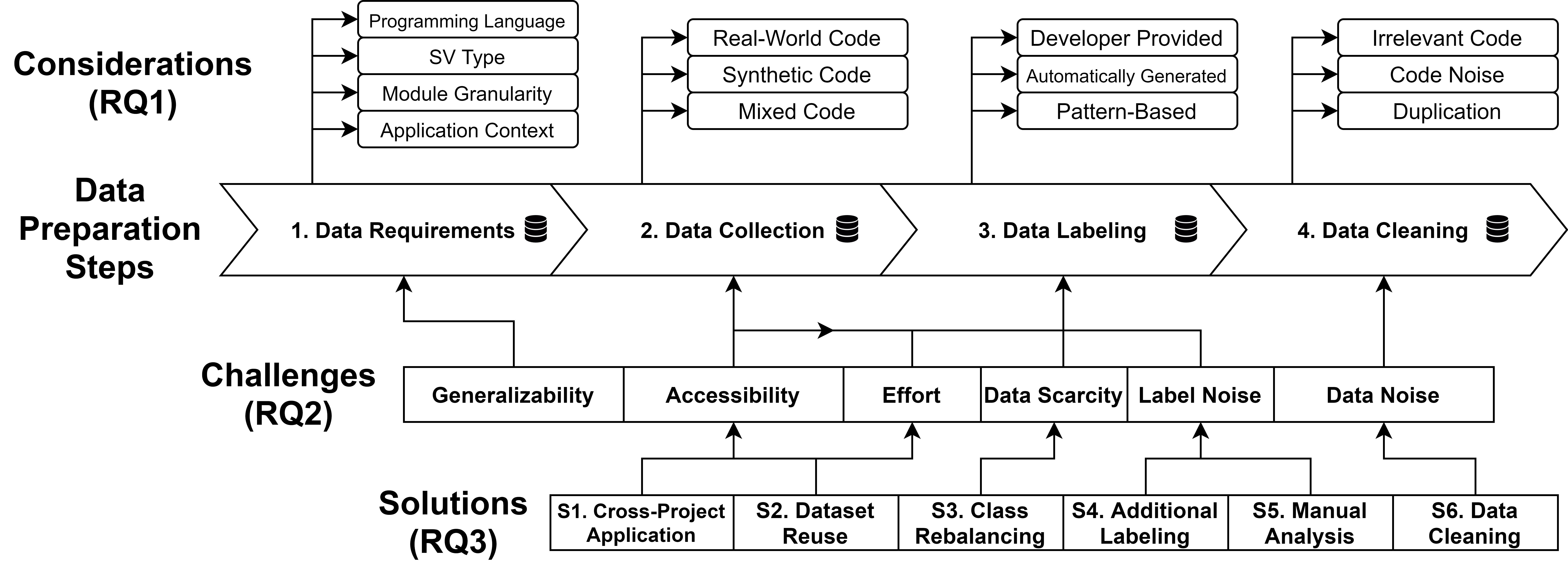}
  \caption{An overview of the SVP considerations, challenges and solutions for data preparation.}
  \label{fig:overview}
\end{figure*}

Based on the findings of this review, we have drawn some actionable recommendations for researchers and practitioners. In order to place our recommendations in the context of the findings from this review, Figure \ref{fig:overview} presents an overview of the findings to our three RQs. Figure \ref{fig:overview} shows the considerations, challenges and current solutions for the data preparation steps of SVP in order to help readers to better understand and interpret our recommendations in this section.  

The validity of SVP research largely relies on the quality of the data used to construct an SVP model. However, as identified in Section \ref{sec:data_challenges}, researchers have faced a plethora of data preparation challenges that may lead to serious pitfalls that should be avoided while training SVP models. 

Due to the current shortcomings of SV datasets and preparation processes, there is an important need for evidence-based research of the existing practices, challenges and solutions in order to advance the state of the art and the state of the practice. Predominantly, there is a need for advancing techniques of real-world data labeling, or creation of more realistic synthetic datasets. Whilst there are several research studies on various related sub-fields, such as fault localization~\cite{zhou2012should}, bug classification~\cite{jiang2020ltrwes}, bug seeding~\cite{jia2010analysis}, or crowd testing~\cite{leicht2017leveraging}, their findings have yet to be leveraged for providing reliable solutions to to address the open research problems. Our identification of the existing obstacles to achieving high quality data for SVP is expected to help researchers to develop suitable methods for overcoming the current barriers for improvement. 

Several studies have called for a need of a \textit{gold-standard} dataset~\cite{hanif2021rise, zeng2020software, lin2020software}; one that overcomes the identified data preparation challenges. We have identified several areas of data preparation solutions in Section \ref{sec:data_solutions} that researchers have used to help address the data preparation challenges. Whilst we recommend that researchers and practitioners also adopt these solutions as they provide an initial foundation for data quality, these solutions are, by no means, complete to produce a gold-standard dataset. Particularly, many challenges for Generalizability remain unaddressed, and the solutions for Label Noise are insufficient. Manual analysis (\textbf{S5}) is often infeasible or inaccurate, and additional labeling (\textbf{S6}) can add additional noise or inconsistencies into a dataset. Furthermore, SVP research has operated at a variety of different contexts, which makes the use of standardized datasets difficult. Different researchers may favor different levels of granularity or applications domains. 

That is why rather than proposing the guidelines for developing a gold-standard dataset, we instead propose a set of recommendations for the future research directions that are expected to help improve the SV data preparation processes by addressing the identified data challenges. Although there are many potential recommendations that can be made, we have formed our discussion from actionable observations obtained from the primary studies and supporting literature, that have the potential of delivering immediate benefit to SVP researchers and practitioners.

\textbf{R1. Consideration of label noise in the negative class.} Fully supervised learning for SVP requires a positive class (vulnerable modules) and a negative class (non-vulnerable modules). The quality of both of these classes is highly important to ensure that a model is able to learn the appropriate patterns in the data. Whilst much efforts and manual inspection have been expended to ensure the quality of the vulnerable labels [P1, P2, P4, P7, P8, P10, P11, P14, P25, P26, P28, P29, P30, P31, P32, P42, P44, P35, P46], equal efforts have not been expended for the non-vulnerable labels. The most commonly used assumption is to simply take all remaining unlabeled files to form the negative class. However, due to incomplete reporting (\textbf{Ch11}), we have observed that there can be considerable inaccuracies in the non-vulnerable class; researchers have only been able to obtain a data source for the vulnerable labels (i.e., the bug reports), but there is no source of labels for non-vulnerable data.

The unavailability of labels for non-vulnerable data adds inherent label noise to the non-vulnerable labels of SV datasets. Such label noise would affect the reliability and trustworthiness of the produced SVP models, as incorrect patterns may be inferred from the mislabeled instances. Hence, there should be more consideration of additional broader label sources to reduce label noise. With proper consideration, additional label sources such as static analysis [P37, P46] can help uncover mislabeled vulnerabilities. Zheng et al. \cite{zheng2021d2a} used differential analysis to increase the confidence of static analysis labels for vulnerability fixes. Crowd-testing~\cite{leicht2017leveraging} may be another potential label source that can minimize individual effort; crowdsourced labels are already used for a variety of benchmark datasets in the ML domain~\cite{tsipras2020imagenet}. Producing more vulnerable labels would also assist in dealing with the challenges in Data Scarcity (\textbf{Ch9}~\&~\textbf{10}). 

However, ensuring the absence of vulnerabilities from code modules is difficult \cite{weinberg2008perfect}. As high-quality real world datasets are currently unavailable, we also suggest a relaxation of labeling and fully-supervised learning requirements. The negative class may be more accurately considered as an unlabeled set. In this scenario, semi-supervised methods, such as self-training or Positive and Unlabeled (PU) learning can be applied to help uncover knowledge from the unlabeled set~\cite{le2020puminer}. 

\textbf{R2. Consideration of timeliness.} Timeliness describes the temporal aspects of data \cite{sidi2012data}. There are two main components for timeliness. The first is the currency of data; the age of data in use compared to when it was collected. Practitioners ought to generally avoid using out-of-date data as it can lose relevancy to contemporary settings. This is particularly an issue for SV data, as vulnerabilities take time to be discovered (\textbf{Ch11}), hence, delayed data collection effort can usually obtain more complete information \cite{jimenez2019importance}. For instance, 10 vulnerabilities may have been reported for a codebase after three months, but an additional 10 vulnerabilities may have been reported one year later. Hence, a lack of consideration for data timeliness will lead to mislabelled instances and a lack of completeness. Researchers ought to develop methods for updating SV datasets to ensure they use more current label data and increase the number of positive samples. This will ensure the datasets are more complete upon use, and consequently allow for more reliable SVP models. 

The second aspect of timeliness relates to the temporal nature of the data accumulation; historical artefacts are created incrementally over the lifecycle of a project. This poses an issue for SV data due to concept drift; the vulnerability data and patterns change over time with the emergence of new concepts~\cite{le2019automated}. Failure to account for this temporal nature, that is, preserving data order for model training and validation, has been shown to produce unreliable models and impact the real-world generalizability~\cite{mcintosh2017fix, falessi2020need}. However, only 13\% of the reviewed studies (8 out of 61) considered a time-based ordering of the data. Hence, there needs to be further investigations into concept drift and use of time-based ordering in order to produce more reliable SVP models. 

\textbf{R3. Use of data visualization.} Data understandability describes the ability to comprehend data~\cite{pipino2002data}. Exploratory data analysis is an integral part of data science, as it helps practitioners to understand data quality issues, imbalances, and relationships within data. This is particularly important for ML as it allows practitioners to identify the necessary data cleaning practices and to conduct feature engineering~\cite{mitchell1999machine}. SV data can be complex due to the intricacies of the code and security weaknesses. Despite this, not much effort has been reported by researchers into data understanding and visualization. 

Data visualization is one of the most powerful techniques for data understandability~\cite{mitchell1999machine}, but only a few of the reviewed studies did visual exploration of their data. Neuhaus and colleagues [P13] visualized the locations of SVs in a codebase and observed that the distribution of SVs are scarce and irregular. This motivated their search for specific code patterns that can be used to describe the irregular distribution. Chowdhury and Zulkerine [P19] used data visualization to identify that files are unlikely to contain SVs in multiple different versions; hence, they determined that vulnerability history of a module may be a poor indicator. The future efforts in data understandability for SV datasets, using similar techniques to the aforementioned ones, may be able to yield novel impactful insights for this data, and assist in SVP model creation. Better data understandability can also help practitioners who lack data science expertise with setting up SVP models. 

\textbf{R4. Creation and use of diverse language datasets.} We observe in Figure \ref{fig:datasets_in_languages} that researchers have primarily constructed datasets for singular languages, as the semantic and syntactic differences of programming languages creates difficulties towards applying models across languages. However, single-language datasets pose inherent coverage problems (\textbf{Ch3}) as modern software development is typically not conducted using a single language \cite{kochhar2016}. SVs can also be present in cross-language operations that use foreign function interfaces. Furthermore, as dataset creation is expensive (\textbf{Ch7}), the lack of multi-language datasets creates scalability problems if practitioners intend to add or switch programming languages. Existing SV datasets have only been constructed for popular languages such as C/C++, Java or PHP, which makes application of SVP models difficult to more obscure languages. Hence, there is need for techniques that can efficiently create or utilise diverse language datasets. Our review found only one study [P53] investigated a cross-language approach. 

Software engineering researchers need to increase their efforts for developing datasets supporting cross-language analysis by leveraging the outcomes of the research in the related areas of software engineering. For example, multilingual source code analysis is an emerging research domain \cite{mushtaq2017multilingual}, that has seen success predominantly in static analysis. Bug seeding is another relevant research field that intends to import bugs from the existing projects to new ones~\cite{jia2010analysis}. We speculate that bug seeding can be used to import bugs from one language into a dataset of a semantically similar language, hence, allowing efficient dataset creation of obscure languages. SVP research will benefit from investigation and adoption of these techniques. More diverse language datasets will increase the completeness and scalability of SVP models for practitioners. 

\textbf{R5. Use of data quality assessment criteria.} We observed in \textbf{Ch13} that the quality of most SVP datasets is unknown and even unconsidered. If researchers and practitioners are able to determine the quality of their datasets, they are expected to make informed decisions about the validity and reliability of the constructed SVP models. Data quality is an inherent requirement of most ML-based systems~\cite{amershi2019software}. Hence, data quality assessment criteria ought to be developed to assess SV datasets in order to enable quantifiable analysis and verification of data quality issues. 

Whilst data quality assessment is a common practice for organisations and data-centric industries~\cite{pipino2002data, sidi2012data}, general data quality dimensions may not be directly applicable or measurable for SV datasets. These generalised data quality patterns need to be customized for SVP. Specific data quality assessment criteria can be determined through requirements-driven approaches \cite{zhang2019discovering}. Our categorisation of SV data challenges in Table \ref{tab:data_challenges} is expected to potentially help in identifying the relevant data quality dimensions and requirements. 

\textbf{R6. Better data sharing and governance.} Although data sharing and reuse have been popular for SVP (\textbf{S2}), we observed that there are many issues regarding the availability of these datasets. Furthermore, the incomplete reporting practices of SV data has led to a lack of trustworthiness. The reporting for the datasets has often been insufficient to allow for proper replication, making the datasets hard to validate. For instance, many of the reviewed studies did not report the version of the data source or specific extraction steps. Furthermore, most of the data preparation processes use extensive manual inspection or labeling (\textbf{S5}), which is hard to replicate due to its subjectivity. Whilst some of the studies have attempted to address this problem by making their datasets available, the code and methods used to create these datasets have been rarely shared; this situation has made the reproduction and adaptation efforts very difficult. For example, Riom and colleagues ~\cite{riom2021revisiting} found the replication of a seminal work [P16] infeasible due to these challenges. 

We observe that the underlying problem of the abovementioned challenges is a lack of proper data reporting and storage efforts. Hence, it can be recommended that researchers make more efforts in the future to specify the exact details and processes of data preparation in order to improve data reuse and augmentation. Gebru and colleagues ~\cite{gebru2018datasheets} propose a process called \textit{Datasheets for Datasets}. Such processes of dataset documentation promises to also help practitioners to better understand the capabilities and implications of the produced SVP models. 

Another potential solution is open data sharing platforms or repositories for enhancing data availability. Such platforms can also provide vital information regarding data provenance and quality. These efforts will potentially also encourage data maintenance, which is another issue that we have discussed in \textbf{R2}. Considerable efforts towards such a platform have already been made in the Software Engineering domain through the PROMISE repository~\cite{promise2005} and the SEACRAFT repository~\cite{seacraftrepo}. However, these efforts still currently fail to ensure data provenance and maintenance. 

Ultimately there is a need for better industrial and corporate engagement with public datasets. Direct industrial collaboration often leads to higher impact software engineering research \cite{garousi2016challenges}. Such involvement will provide more realistic, complete and trustworthy datasets, due to increased data provenance and accessibility. This cooperation would also allow SVP evaluation to extend beyond open-source repositories to private-source code. As technical advancements continue to be made for SVP models, researchers should move towards industrial demonstrations and case studies that can encourage further corporate participation and application. Garousi et al. \cite{garousi2016challenges} have outlined several best practices towards ensuring effective industry-academia collaboration. 

\section{Threats to Validity}
\label{sec:threats}

This review has been designed and executed by carefully following the guidelines for SLR provided by Kitchenham et al.~\cite{kitchenham2004procedures}. Hence, we have identified the potential validity threats to this SLR and taken appropriate steps to minimize the potential impact of the identified threats as per the SLR guidelines. We discuss the validity threats considered for this SLR below. 

A standard threat to any SLR is selection bias; some relevant papers may be missed during the selection process of the SLR. Whilst there is a possibility that our search and study selection process may have missed some relevant papers, we systematically drove the paper selection process by following the SLR guidelines and the recommended practices to minimize such possibility of missing the relevant papers. For example, we chose a meta search engine, SCOPUS, that indexes all the well-known computer science and software engineering digital libraries such as IEEE, ACM, Elsevier, and Springer. Furthermore, we iteratively refined our search string using the quasi-gold sensitivity approach defined by Zhang et al.~\cite{zhang2011identifying} until we were confident that our search string had retrieved the majority of the key papers in this research area. Finally, we used backward and forward snowballing to help capture any studies that might have been missed during our automatic search. To avoid the study selection bias by the authors, we initially conducted a collaborative pilot study selection on 100 papers to ensure consistency. Furthermore, any paper that an individual author was not confident about including/excluding was discussed between the first two authors before making a final decision. 

Additional validity threats can be introduced through the quality assessment, data extraction and thematic analysis processes of this SLR. Inaccuracies in these processes can be introduced by human-error and researcher-bias. The first two authors jointly carried out the pilot activities for these processes to help ensure consistency. Furthermore, all the data-extraction activities were cross-checked by the authors, with disagreements resolved through discussions. 

Finally, our results may be affected by publication bias; research is biased towards the publication of positive results over negative results~\cite{kitchenham2004procedures}. Hence, it is expected that researchers would also be reluctant to report the major (data) limitations if it is not the focus of a study. As our findings are grounded in the data extracted from the primary studies, we are only able to report the considerations and challenges explicitly discussed in the papers. Hence, our findings may not be exhaustive. However, we assume that our findings are able to capture the major data challenges and considerations, due to the quality and integrity of our selected primary studies which we ensured through rigorous use of our inclusion/exclusion and data quality assessment criteria from Section \ref{sec:selection}. 

\section{Conclusion}
\label{sec:conclusion}
Software Vulnerability Prediction (SVP) approaches have gained significant attention of the software engineering community for ensuring software security. Whilst researchers have developed and evaluated a large number of SVP approaches, they have also identified several challenges that need to be addressed for wider adoption of the reported approaches. Given the importance of the topic and the amount of available literature that is largely dispersed, it was needed and timely to invest efforts in a Systemization of Knowledge (SoK) of the available peer-reviewed literature in order to highlight the key challenges of data preparation phases for developing, evaluating, and deploying SVP approaches. 

We have carried out a systematic literature review of 61 papers by following the well-known SLR guidelines. The main aim of our study was to identify the considerations, challenges and solutions of data preparation for SVP, by answering three research questions. Firstly, we have identified the major decisions made by researchers for the data preparation processes, to help inform the state of the practice. Secondly, through our thematic analysis, we have derived a taxonomy of 16 identified data challenges that researchers face for constructing data-driven methods for SVP. These challenges involve the data Generalizability, Accessibility, label collection Effort and Scarcity, and both Label and Data Noise. Thirdly, we also categorized and mapped the data preparation related solutions reported in the reviewed papers in order to highlight their understanding and utility. 

We have found that the identified challenges are particularly pertinent to the data labelling process that has consequently attracted the majority of the identified solutions. Due to these significant challenges, data reuse is a common solution to reduce data construction effort and difficulties. Alternatively, the use of synthetic datasets is an attractive option as these datasets artificially solve challenges for Accessibility, Effort, Data Scarcity and Label Noise, but these datasets experience significant challenges with Generalizability, which severely limits their value. 

We assert that the findings of our SLR will help researchers and practitioners to understand the key SVP data preparation considerations and challenges. By consolidating the state of the practice into an integrated source of information, SoK, this study is expected to assist practitioners in improving their data preparation practices for building and deploying SVP models. Furthermore, we believe that the reported taxonomy of the data preparation challenges will be used to identify and classify the data preparation for SVP related challenges that practitioners may encounter; and our categorization of the identified solutions is expected to help identify the best practices of data preparation for SVP models. Such improvements are expected to improve the quality of SVP models, as their efficacy hinges on the data quality; \textit{Garbage In, Garbage Out}. 

The findings of this review will inform the future research that is expected to address the challenges for which the reviewed studies do not provide appropriate solutions. We have derived six recommendations from the findings of this review for providing actionable suggestions to SVP researchers and practitioners. Whilst we acknowledge that our study does not provide a complete overview of the SVP process on its own, the data is undoubtedly one of the most important components for any data-driven process; hence, we take one of the first steps to highlight this area where the future research efforts can make significant advances in the data preparation and data quality state-of-the-art. Such advances will enable the creation and use of more reliable and trustworthy SVP approaches supporting automated software security analytics. 

\ifCLASSOPTIONcompsoc
  \section*{Acknowledgments}
\else
  \section*{Acknowledgment}
\fi

This work has been supported by the Cyber Security Cooperative Research Centre Limited whose activities are partially funded by the Australian Government’s Cooperative Research Centre Programme. 


%

\appendix[Primary Studies]
\label{appendix}
\begin{footnotesize}
\begin{enumerate}[leftmargin=0.5cm]
	\item[P1] J. Walden, J. Stuckman and R. Scandariato, "Predicting vulnerable components: Software metrics vs text mining," \textit{International Symposium on Software Reliability Engineering}, 2014.
	\item[P2] Y. Zhou, S. Liu, J. Siow, X. Du and Y. Liu, "Devign: Effective vulnerability identification by learning comprehensive program semantics via graph neural networks," \textit{Conference on Neural Information Processing Systems}, 2019.
	\item[P3] J. Tian, W. Xing and Z. Li, "BVDetector: A program slice-based binary code vulnerability intelligent detection system," \textit{Information and Software Technology}, 2020.
	\item[P4] C. Theisen and L. Williams, "Better together: Comparing vulnerability prediction models," \textit{Information and Software Technology}, 2020.
	\item[P5] P. Morrison, K. Herzig, B. Murphy and L. Williams, "Challenges with applying vulnerability prediction models," \textit{Symposium and Bootcamp on the Science of Security}, 2015.
	\item[P6] A. Fidalgo, I. Medeiros, P. Antunes and N. Neves, "Towards a Deep Learning Model for Vulnerability Detection on Web Application Variants," \textit{International Conference on Software Testing, Verification and Validation Workshops}, 2020.
	\item[P7] X. Ban, S. Liu, C. Chen and C. Chua, "A performance evaluation of deep-learnt features for software vulnerability detection," \textit{Concurrency and Computation: Practice and Experience}, 2019.
	\item[P8] Z. Li, D. Zou, J. Tang, Z. Zhang, M. Sun and H. Jin, "A comparative study of deep learning-based vulnerability detection system," \textit{IEEE Access}, 2019.
	\item[P9] T. Nguyen, T. Le, K. Nguyen, O. de Vel, P. Montague, J. Grundy and D. Phung, "Deep Cost-Sensitive Kernel Machine for Binary Software Vulnerability Detection," \textit{Pacific-Asia Conference on Knowledge Discovery and Data Mining}, 2020.
	\item[P10] G. Lin, J. Zhang, W. Luo, L. Pan, O. D. Vel, P. Montague and Y. Xiang, "Software Vulnerability Discovery via Learning Multi-domain Knowledge Bases," \textit{IEEE Transactions on Dependable and Secure Computing}, 2019.
	\item[P11] M. Gegick, L. Williams, J. Osborne and M. Vouk, "Prioritizing software security fortification through code-level metrics," \textit{ACM Workshop on Quality of protection}, 2008.
	\item[P12] M. Jimenez, M. Papadakis and Y. L. Traon, "Vulnerability prediction models: A case study on the Linux Kernel," \textit{International Working Conference on Source Code Analysis and Manipulation}, 2016.
	\item[P13] S. Neuhaus, T. Zimmermann, C. Holler and A. Zeller, "Predicting vulnerable software components," \textit{ACM Conference on Computer and Communications Security}, 2007.
	\item[P14] S. Liu, G. Lin, Q. Han, S. Wen, J. Zhang and Y. Xiang, "DeepBalance: Deep-Learning and Fuzzy Oversampling for Vulnerability Detection," \textit{IEEE Transactions on Fuzzy Systems}, 2020.
	\item[P15] M. Zagane, M. K. Abdi and M. Alenezi, "Deep Learning for Software Vulnerabilities Detection Using Code Metrics," \textit{IEEE Access}, 2020.
	\item[P16] H. Perl, S. Dechand, M. Smith, D. Arp, F. Yamaguchi, K. Rieck, S. Fahl and Y. Acar, "VCCFinder: Finding potential vulnerabilities in open-source projects to assist code audits," \textit{ACM SIGSAC Conference on Computer and Communications Security}, 2015.
	\item[P17] Y. Joonseok, R. Duksan and B. Jongmoon, "Improving vulnerability prediction accuracy with Secure Coding Standard violation measures," \textit{International Conference on Big Data and Smart Computing (BigComp)}, 2016.
	\item[P18] V. Nguyen, T. Le, T. Le, K. Nguyen, O. De Vel, P. Montague, L. Qu and D. Phung, "Deep Domain Adaptation for Vulnerable Code Function Identification," \textit{International Joint Conference on Neural Networks}, 2019.
	\item[P19] I. Chowdhury and M. Zulkernine, "Using complexity, coupling, and cohesion metrics as early indicators of vulnerabilities," \textit{Journal of Systems Architecture}, 2011.
	\item[P20] H. K. Dam, T. Tran, T. Pham, S. W. Ng, J. Grundy and A. Ghose, "Automatic Feature Learning for Predicting Vulnerable Software Components," \textit{IEEE Transactions on Software Engineering}, 2018.
	\item[P21] A. Meneely and L. Williams, "Strengthening the empirical analysis of the relationship between Linus' Law and software security," \textit{International Symposium on Empirical Software Engineering and Measurement}, 2010.
	\item[P22] X. Chen, Y. Zhao, Z. Cui, G. Meng, Y. Liu and Z. Wang, "Large-Scale Empirical Studies on Effort-Aware Security Vulnerability Prediction Methods," \textit{IEEE Transactions on Reliability}, 2020.
	\item[P23] X. Cheng, H. Wang, J. Hua, M. Zhang, G. Xu, L. Yi and Y. Sui, "Static detection of control-flow-related vulnerabilities using graph embedding," \textit{International Conference on Engineering of Complex Computer Systems}, 2019.
	\item[P24] V. H. Nguyen and L. M. S. Tran, "Predicting vulnerable software components with dependency graphs," \textit{International Workshop on Security Measurements and Metrics}, 2010.
	\item[P25] L. K. Shar and H. B. K. Tan, "Predicting SQL injection and cross site scripting vulnerabilities through mining input sanitization patterns," \textit{Information and Software Technology}, 2013.
	\item[P26] S. Liu, G. Lin, L. Qu, J. Zhang, O. D. Vel, P. Montague and Y. Xiang, "CD-VulD: Cross-Domain Vulnerability Discovery based on Deep Domain Adaptation," \textit{IEEE Transactions on Dependable and Secure Computing}, 2020.
	\item[P27] B. M. Padmanabhuni and H. B. K. Tan, "Buffer Overflow Vulnerability Prediction from x86 Executables Using Static Analysis and Machine Learning," \textit{Annual Computer Software and Applications Conference}, 2015.
	\item[P28] W. Zheng, J. Gao, X. Wu, F. Liu, Y. Xun, G. Liu and X. Chen, "The impact factors on the performance of machine learning-based vulnerability detection: A comparative study," \textit{Journal of Systems and Software}, 2020.
	\item[P29] H. Wang, G. Ye, Z. Tang, S. H. Tan, S. Huang, D. Fang, Y. Feng, L. Bian and Z. Wang, "Combining Graph-Based Learning With Automated Data Collection for Code Vulnerability Detection," \textit{IEEE Transactions on Information Forensics and Security}, 2021.
	\item[P30] Y. Shin, A. Meneely, L. Williams and J. A. Osborne, "Evaluating complexity, code churn, and developer activity metrics as indicators of software vulnerabilities," \textit{IEEE Transactions on Software Engineering}, 2011.
	\item[P31] G. Lin, J. Zhang, W. Luo, L. Pan, Y. Xiang, O. De Vel and P. Montague, "Cross-Project Transfer Representation Learning for Vulnerable Function Discovery," \textit{IEEE Transactions on Industrial Informatics}, 2018.
	\item[P32] Z. Li, D. Zou, S. Xu, X. Ou, H. Jin, S. Wang, Z. Deng and Y. Zhong, "Vuldeepecker: A deep learning-based system for vulnerability detection," \textit{Usenix Network and Distributed System Security Symposium}, 2018.
	\item[P33] Y. Shin and L. Williams, "Can traditional fault prediction models be used for vulnerability prediction?," \textit{Empirical Software Engineering}, 2013.
	\item[P34] N. Saccente, J. Dehlinger, L. Deng, S. Chakraborty and Y. Xiong, "Project achilles: A prototype tool for static method-level vulnerability detection of Java source code using a recurrent neural network," \textit{International Conference on Automated Software Engineering Workshop}, 2019.
	\item[P35] Y. Zhang, D. Lo, X. Xia, B. Xu, J. Sun and S. Li, "Combining Software Metrics and Text Features for Vulnerable File Prediction," \textit{International Conference on Engineering of Complex Computer Systems}, 2016.
	\item[P36] M. Jimenez, R. Rwemalika, M. Papadakis, F. Sarro, Y. Le Traon and M. Harman, "The importance of accounting for real-world labeling when predicting software vulnerabilities," \textit{ACM Joint Meeting on European Software Engineering Conference and Symposium on the Foundations of Software Engineering}, 2019.
	\item[P37] S. Moshtari, A. Sami and M. Azimi, "Using complexity metrics to improve software security," \textit{Computer Fraud, Security}, 2013.
	\item[P38] D. Zou, S. Wang, S. Xu, Z. Li and H. Jin, "$\mu$VulDeePecker: A Deep Learning-Based System for Multiclass Vulnerability Detection," \textit{IEEE Transactions on Dependable and Secure Computing}, 2019.
	\item[P39] S. Ghaffarian and H. Shahriari, "Neural software vulnerability analysis using rich intermediate graph representations of programs," \textit{Information Sciences}, 2021.
	\item[P40] M. Siavvas, D. Tsoukalas, M. Jankovic, D. Kehagias and D. Tzovaras, "Technical debt as an indicator of software security risk: a machine learning approach for software development enterprises," \textit{Enterprise Information Systems}, 2020.
	\item[P41] I. Abunadi and M. Alenezi, "An empirical investigation of security vulnerabilities within web applications," \textit{Journal of Universal Computer Science}, 2016.
	\item[P42] L. Yang, X. Li and Y. Yu, "VulDigger: A Just-in-Time and Cost-Aware Tool for Digging Vulnerability-Contributing Changes," \textit{IEEE Global Communications Conference}, 2017.
	\item[P43] V. Nguyen, T. Le, O. de Vel, P. Montague, J. Grundy and D. Phung, "Dual-Component Deep Domain Adaptation: A New Approach for Cross Project Software Vulnerability Detection," \textit{Pacific-Asia Conference on Knowledge Discovery and Data Mining}, 2020.
	\item[P44] Z. Li, D. Zou, S. Xu, H. Jin, Y. Zhu and Z. Chen, "SySeVR: A Framework for Using Deep Learning to Detect Software Vulnerabilities," \textit{IEEE Transactions on Dependable and Secure Computing}, 2021.
	\item[P45] S. Liu, M. Dibaei, Y. Tai, C. Chen, J. Zhang and Y. Xiang, "Cyber Vulnerability Intelligence for Internet of Things Binary," \textit{IEEE Transactions on Industrial Informatics}, 2020.
	\item[P46] R. Scandariato, J. Walden, A. Hovsepyan and W. Joosen, "Predicting vulnerable software components via text mining," \textit{IEEE Transactions on Software Engineering}, 2014.
	\item[P47] N. Medeiros, N. Ivaki, P. Costa and M. Vieira, "Software metrics as indicators of security vulnerabilities," \textit{Software Reliability Engineering}, 2017.
	\item[P48] M. A. Albahar, "A Modified Maximal Divergence Sequential Auto-Encoder and Time Delay Neural Network Models for Vulnerable Binary Codes Detection," \textit{IEEE Access}, 2020.
	\item[P49] X. Duan, J. Wu, S. Ji, Z. Rui, T. Luo, M. Yang and Y. Wu, "Vulsniper: Focus your attention to shoot fine-grained vulnerabilities," \textit{International Joint Conference on Artificial Intelligence}, 2019.
	\item[P50] Y. Fang, S. Han, C. Huang and R. Wu, "TAP: A static analysis model for PHP vulnerabilities based on token and deep learning technology," \textit{PloS one}, 2019.
	\item[P51] D. Cao, J. Huang, X. Zhang and X. Liu, "FTCLNet: Convolutional LSTM with Fourier Transform for Vulnerability Detection," \textit{Trust, Security and Privacy in Computing and Communications}, 2020.
	\item[P52] Z. Yu, C. Theisen, L. Williams and T. Menzies, "Improving Vulnerability Inspection Efficiency Using Active Learning," \textit{IEEE Transactions on Software Engineering}, 2019.
	\item[P53] A. Figueiredo, T. Lide, D. Matos and M. Correia, "MERLIN: Multi-Language Web Vulnerability Detection," \textit{International Symposium on Network Computing and Applications}, 2020.
	\item[P54] Z. Bilgin, M. A. Ersoy, E. U. Soykan, E. Tomur, P. Comak and L. Karacay, "Vulnerability Prediction from Source Code Using Machine Learning," \textit{IEEE Access}, 2020.
	\item[P55] X. Chen, Z. Yuan, Z. Cui, D. Zhang and X. Ju, "Empirical studies on the impact of filter‐based ranking feature selection on security vulnerability prediction," \textit{IET Software}, 2020.
	\item[P56] J. Stuckman, J. Walden and R. Scandariato, "The Effect of Dimensionality Reduction on Software Vulnerability Prediction Models," \textit{IEEE Transactions on Reliability}, 2017.
	\item[P57] Y. Tang, F. Zhao, Y. Yang, H. Lu, Y. Zhou and B. Xu, "Predicting Vulnerable Components via Text Mining or Software Metrics? An Effort-Aware Perspective," \textit{IEEE International Conference on Software Quality, Reliability and Security}, 2015.
	\item[P58] J. Domanska, M. Siavvas and E. Gelenbe, "Efficient Feature Selection for Static Analysis Vulnerability Prediction," \textit{K. Filus, P. Boryszko, Sensors}, 2021.
	\item[P59] Y. Shin and L. Williams, "An initial study on the use of execution complexity metrics as indicators of software vulnerabilities," \textit{International Workshop on Software Engineering for Secure Systems}, 2011.
	\item[P60] N. Medeiros, N. Ivaki, P. Costa and M. Vieira, "Vulnerable Code Detection Using Software Metrics and Machine Learning," \textit{IEEE Access}, 2020.
	\item[P61] R. Li, C. Feng, X. Zhang and C. Tang, "A Lightweight Assisted Vulnerability Discovery Method Using Deep Neural Networks," \textit{IEEE Access}, 2019.
\end{enumerate}
\end{footnotesize}

\ifCLASSOPTIONcaptionsoff
  \newpage
\fi



%
\bibliographystyle{IEEEtran}
\bibliography{bibfile}

%








\end{document}